\documentclass[manuscript]{rmaa}

\usepackage{paralist}

\usepackage{psfrag,color}

\usepackage[utf8]{inputenc}
\usepackage[T1]{fontenc}

\usepackage{multirow}
\usepackage{longtable}
\usepackage{subcaption}
\usepackage{subfig}
\usepackage{graphicx}
\usepackage{booktabs}
\usepackage{xcolor}
\usepackage{comment}
\usepackage{arydshln}

\newcommand\degree{\hbox{$^\circ$}}

\title{A First Quantitative Characterization of Colombian Night Sky Through All-Sky Photometry} 

\author{
J. P. Uchima-Tamayo\altaffilmark{1}, 
R. Angeloni\altaffilmark{2},
M. Jaque Arancibia \altaffilmark{1,3},
C. Goez Theran\altaffilmark{4} and
J. F. R\'ua Restrepo\altaffilmark{5}}

\altaffiltext{1}{Departamento de Astronom\'ia, Universidad de La Serena, La Serena, Chile (corresponding author: juan.uchima@userena.cl).}

\altaffiltext{2}{Gemini Observatory, NSF's NOIRLab, La Serena, Chile.}

\altaffiltext{3}{Instituto Multidisciplinario de Investigaci\'on y Postgrado, Universidad de La Serena, La Serena, Chile}

\altaffiltext{4}{Oficina de Olimpiadas Colombianas, Universidad Antonio Nari\~no, Bogot\'a, Colombia}

\altaffiltext{5}{Observatorio Astrosur, Villavieja, Colombia}

\shortauthor{Uchima-Tamayo et al.}
\shorttitle{Characterization of Colombian Night Sky}

\fulladdresses{
\item Rodolfo Angeloni: Gemini Observatory, NSF's NOIRLab, Av.~J.~Cisternas 1500 N, 1720236, La Serena, Chile (rodolfo.angeloni@noirlab.edu).
\item Cristian Goez Theran: Oficina de Olimpiadas Colombianas, Universidad Antonio Nari\~no, Calle 58a \# 37 - 94, Bogot\'a, Colombia (cristian.goez@uan.edu.co).

\item Marcelo Jaque Arancibia \& Juan Pablo Uchima-Tamayo: Departamento de Astronom\'ia, Universidad de La Serena,   Av.~R.~Bitrán 1305, La Serena, Chile (marcelo.jaque, juan.uchima@userena.cl).

\item Marcelo Jaque Arancibia: Instituto Multidisciplinario de Investigaci\'on y Postgrado, Universidad de La Serena, Av.~R.~Bitrán 1305, La Serena, Chile.

\item Javier Fernando R\'ua Restrepo: Observatorio Astrosur, Km 8 desierto de la Tatacoa, Villavieja, Colombia (astrosur@yahoo.com).
}

\listofauthors{J. P. Uchima-Tamayo, R. Angeloni, M. Jaque Arancibia, C. Goez Theran \& J. F. R\'ua Restrepo}
\indexauthor{Uchima-Tamayo, J. P.}
\indexauthor{Angeloni, R.}
\indexauthor{Jaque Arancibia, M}
\indexauthor{Goez Theran, C}
\indexauthor{R\'ua Restrepo, J. F.}

\abstract{Light pollution, a rapidly escalating anthropogenic phenomenon driven by the excessive and often inefficient use of artificial lighting, has profound implications for astronomy, ecology, and human health. This study presents the first comprehensive characterization of night sky quality in Colombia, focusing on sites of astronomical and ecological significance. The selected locations include the Astronomical Observatory of UTP, the Tatacoa Desert, the Bogot\'a Botanical Garden, and Cerro Guadalupe. Utilizing the Sky Quality Camera, we collected all-sky data to measure surface brightness and correlated color temperature of the night sky. Our findings reveal a significant loss of natural sky visibility in urban areas and demonstrate the detrimental effects of artificial lighting on critical astronomical sites such as La Tatacoa. This study provides a crucial foundation for future research and informs the development of public policies aimed at preserving the night sky.}

\resumen{La contaminaci\'on lum\'inica, un fen\'omeno antropog\'enico en r\'apido crecimiento, es impulsada por el uso excesivo de tecnolog\'ias de iluminaci\'on. Este mal uso afecta la astronom\'ia, ecolog\'ia y salud humana. En este estudio, caracterizamos por primera vez la calidad del cielo nocturno en Colombia desde sitios de inter\'es astron\'omico y ecol\'ogico. Los lugares analizados incluyen el Observatorio Astron\'omico de la UTP, el desierto de La Tatacoa, el Jard\'in Bot\'anico de Bogot\'a y el Cerro Guadalupe. Utilizando la Sky Quality Camera, obtuvimos datos all-sky sobre el brillo superficial del cielo nocturno y la temperatura de color correlacionada. Los resultados evidencian la p\'erdida de las fuentes naturales del cielo en \'areas urbanas y c\'omo la luz artificial afecta sitios astron\'omicos clave, como La Tatacoa. Este estudio sienta las bases para futuras investigaciones y pol\'iticas p\'ublicas destinadas a preservar el cielo nocturno.}

\addkeyword{Light pollution}
\addkeyword{instrumentation: detectors}
\addkeyword{methods: observational}
\addkeyword{techniques: photometric}
\addkeyword{software: data analysis}

\begin{document}
\maketitle

\section{Introduction} \label{Sec_1}

Light pollution, defined as the inappropriate or excessive use of artificial light, was first described by \citet{walker1973light} and \citet{riegel1973light} as a loss of night sky darkness, causing the loss of the ability to observe celestial objects (\citealt{cinzano2000artificial}; \citealt{falchi2011limiting}). Later, \citet{verheijen1985photopollution} proposed the term \textquotedblleft photopollution\textquotedblright \ to refer to artificial light that adversely affects wildlife. This phenomenon can disrupt the behavioral, feeding, breeding, and migration patterns of various species \citep{longcore2004ecological, gaston2013ecological, davies2014nature, holker202111}. Also, light pollution is a complex disruptor of the nocturnal environment, impacting everything from molecular processes to ecosystems. Its effects can be observed from the immediate vicinity of light sources to hundreds of kilometers away. The impacts of light pollution occur over a wide range of time scales, from seconds or less to decades, all within a general trend of interannual increase \citep{bara2023artificial}.

For nearly two decades, there has been a significant increase in studies examining the effects of artificial light on astronomy and ecology. \citet{holker2010dark} introduced the term Artificial Light At Night (ALAN) to describe techniques for illuminating future landscapes and contributing to environmental protection. ALAN has rapidly evolved into an interdisciplinary field, influencing various aspects including human health (\citealt{cho2015effects};  \citealt{svechkina2020impact}; \citealt{deprato2024influence}), economy (\citealt{gallaway2010economics}; \citealt{mitchell2019dark}), philosophy \citep{henderson2010valuing}, urban planning \citep{zielinska2019urban}, and tourism (\citealt{c2019astrotourism}; \citealt{varela2023increasing}). Despite the progress made in understanding and addressing light pollution, significant challenges persist in mitigating its effects, especially in densely populated urban environments (\citealt{aube2016spectral}; \citealt{kyba2017artificially}).

ALAN corresponds to the artificial contribution to the Night Sky Brightness (NSB). However, for a complete characterization of the sky, it is also necessary to consider the natural contributions, which come from sources not influenced by human activity. They fall into two categories: sources originating in or near the Earth's atmosphere (such as airglow, auroras) and those coming from space beyond the Earth's atmosphere (such as integrated starlight, zodiacal light, galactic light). All these contributions have been described in various studies (\citealt{hanel2018measuring}; \citealt{alarcon2021natural}; \citealt{barentine2022night}). Understanding both artificial and natural contributions is thus essential to accurately assess the impact of light pollution and to develop effective strategies for preserving the night sky \citep{zielinska2020assessment}.

Research conducted by \citet{kyba2017artificially, kyba2023} indicates that over the past decade, the night sky has undergone significant changes due to an increase in the amount of light emitted and the continuous expansion of artificially illuminated areas, contributing to the rise in global skyglow. This has intensified the effects of light pollution, resulting in a much brighter night environment and diminishing the natural darkness of the sky. This shift is primarily driven by the increasing use of more efficient and lower-cost lighting technologies \citep{tsao2010world}. A key technology driving this change is light-emitting diodes (LEDs). Although LEDs are more efficient and durable than traditional lighting technologies, such as high-pressure sodium lamps, they have fundamentally changed the nature of ALAN, despite their advantages in efficiency and durability (\citealt{chang2012light}; \citealt{nair2015perspective}; \citealt{cho2017white}), they have also altered the Spectral Power Distribution (SPD) of ALAN, that is shifting from displaying only a few emission lines to dozens, and even hundreds, of lines \citep{hung2021changes}.

A key factor in assessing the impact of lighting changes is the Correlated Color Temperature (CCT), which quantifies the color of emitted light by comparing it to the light produced by a blackbody at a given temperature, measured in Kelvin (K). The shift in the SPD introduced by LED technology affects not just the intensity, but also the color characteristics of artificial light. This, in turn, influences both the nocturnal environment and biological processes \citep{durmus2022correlated}. Consequently, the transition from traditional lighting to LEDs not only exacerbates light pollution but also alters how we experience and interact with artificial light during nighttime. For example, during the day, the sky typically shows CCT values between 5500 K and 6500 K. At night, these values are largely influenced by artificial lighting, such as city lights. Sodium lamps, for instance, produce lower CCT values, ranging from 2000 K to 3000 K, while white LED lights can generate higher CCT levels, between 4000 K and 6500 K. Furthermore, atmospheric conditions and natural light sources can introduce significant variations in these measurements \citep{jechow2019using}.  Furthermore, CCT has become an important measure due to its considerable impact on people's health and well-being (\citealt{mills2007effect}; \citealt{luo2023effects}). 

To understand and evaluate the impact of ALAN, various measurement techniques have been developed. \citet{mander2023measure} grouped the techniques into four categories.
Spaceborne sensors provide global data on light pollution via satellites. Airborne sensors, including aircraft, balloons and drones, make measurements, allowing for a more accurate assessment of the effects of ALAN on the environment and communities. Ground-based monochromatic sensors, such as the Sky Quality Meter\footnote{http://www.unihedron.com/projects/darksky/} (SQM) described by \citet{cinzano2005night} and the Telescope Encoder and Sky Sensor\footnote{https://tess.stars4all.eu/} (TESS) developed by (\citealt{zamorano2016stars4all}; \citealt{bara2019absolute}), are popular for measuring NSB due to their one-dimensional output and are typically deployed in regions or countries for detailed estimates of the evolution of ALAN (\citealt{posch2018systematic}; \citealt{bertolo2019measurements}).  Additionally, ground-based photometric sensors provide information on the spatial distribution of sky brightness using pseudo-color maps and, when oriented at zenith, typically use a CCD or a CMOS sensor combined with a fast objective lens (\citealt{kollath2010measuring};\citealt{aceituno2011all}). One of the most widely used instruments is the Sky Quality Camera (SQC), designed, developed and sold by Euromix Ltd, Slovenia. It is essentially a calibrated DSLR camera able to provide spatially-resolved NSB (in units of $[mcd/m^2]$ or $[V mag/arcsec^2]$) and CCT (in units of Kelvin [K]) observations.
When pointed towards the zenith, this equipment is able to capture a hemispherical view that includes both the sky and the local horizon. Thanks to is large dynamical range, it has been used under very different skies, from virtually pristine ones down to urban centers heavily light polluted \citep{angeloni2024toward}.
As reported in recent reviews (e.g., \citealt{mander2023measure}), the SQC has today become the reference tool for all-sky photometric studies of light pollution for the optimal balance between cost, ease of use, and richness of information it is able to provide \citep{hanel2018measuring}. It is the instrument also used in this study (Sect.\ref{Sec_2.2}).

In Latin America, studies on light pollution have primarily focused on regions with astronomical infrastructures, such as Chile (\citealt{krisciunas2007optical}; \citealt{muller2011measuring}; \citealt{falchi2023light}; \citealt{angeloni2024toward}), Mexico (\citealt{tovmassian2016astroclimatic}; \citealt{plauchu2017night}), and Argentina (\citealt{aube2014evaluation}; \citealt{iglesias2023daytime}).
In Colombia, however, measurements have been more centered on ecological places, using SQM in locations such as Bogot\'a \citep{aguilera2012evaluacion}, the Bogot\'a Botanical Garden (hereafter BBG - \citealt{urrego2016analisis}), and the Tatacoa Desert (hereafter TD - \citealt{goez2021comparative}). \citet{rueda2023illuminating} presents the most recent study on ALAN in Colombia, examining the expansion of NSB in major cities including Bogot\'a, Medell\'in, Cali, Bucaramanga, Barranquilla, and Cartagena, using Visible Infrared Imaging Radiometer Suite\footnote{https://www.earthdata.nasa.gov/sensors/viirs} (VIIRS) data from 2012 to 2022. Additionally, satellite data have been employed to identify areas with low light pollution and to identify suitable sites for astronomical observation, both in the optical spectrum \citep{arbelaez2020estimating} and in the radio spectrum \citep{molano2017low}.

Colombia is renowned for its megadiverse ecosystems, ranking fourth in plant species richness, fifth in mammals, first in birds, third in reptiles, and second in amphibians, freshwater fish, and butterflies \citep{andrade2011estado}. Despite this notable biodiversity and considering the important impacts of light pollution on ecosystems, there is currently no specific legislation in Colombia to protect fauna from this type of pollution. This lack of regulation is worrying, since recent studies, such as those carried out by \citet{marin2022artificial} and \citet{sanchez2021urbanization}, have shown alterations in the songs of several species of birds due to light pollution in Armenia, a city located within the Humid Forest Biome of the central Andes. These studies highlight the urgent need to implement protection measures to preserve Colombian biodiversity. Additionally, Colombia is recognized as one of the cloudiest regions in the world, largely due to the complex interaction between its geography and atmospheric conditions \citep{poveda2011hydro}. The frequent cloud cover can intensify the effects of light pollution, as artificial light is scattered by the clouds, amplifying its impact. This further disrupts local wildlife and interferes with their natural cycles \citep{kyba2011cloud}.

In contrast to countries such as Spain\footnote{https://www.boe.es/buscar/pdf/2010/BOE-A-2010-20074-consolidado.pdf}, Mexico\footnote{https://www.ensenada.gob.mx/wp-content/uploads/2021/11/Reglamento-para-la-prevencion-de-la-contaminacion-luminica-en-el-Municipio-de-Ensenada-Baja-California.pdf}, and Chile\footnote{https://luminica.mma.gob.cl/norma-luminica/}, which have implemented regulations to control light pollution, Colombia lacks specific legislation aimed at protecting dark skies or properly regulating light pollution. While the country does have technical regulations governing public and general lighting systems, these standards have been criticized as overly lenient and insufficient for preserving the night sky, as noted by \citet{benitez2016eficacia} in a study conducted between 2010 and 2016. Implementing stricter measures is essential to preserve the quality of the night sky. Therefore, society must become aware of and sensitive to the harmful effects of light pollution, promoting sustainable lighting practices \citet{vierdayanti2024promoting}.

This study presents the first comprehensive NSB and CCT maps for the Colombian skies, derived from data collected by SQC during an observing campaign in September 2022. Section \ref{Sec_2} describes the study sites, the data acquisition process, and the data reduction method. Section \ref{Sec_3} presents the results, followed by a discussion in Section \ref{Sec_4}. Finally, Section \ref{Sec_5} outlines the conclusions and final observations.

\section{Materials and Methods} \label{Sec_2}

\subsection{Locations of Interest} \label{Sec_2.1}

\begin{figure*}[!t]
    \centering
    \includegraphics[width=\textwidth]{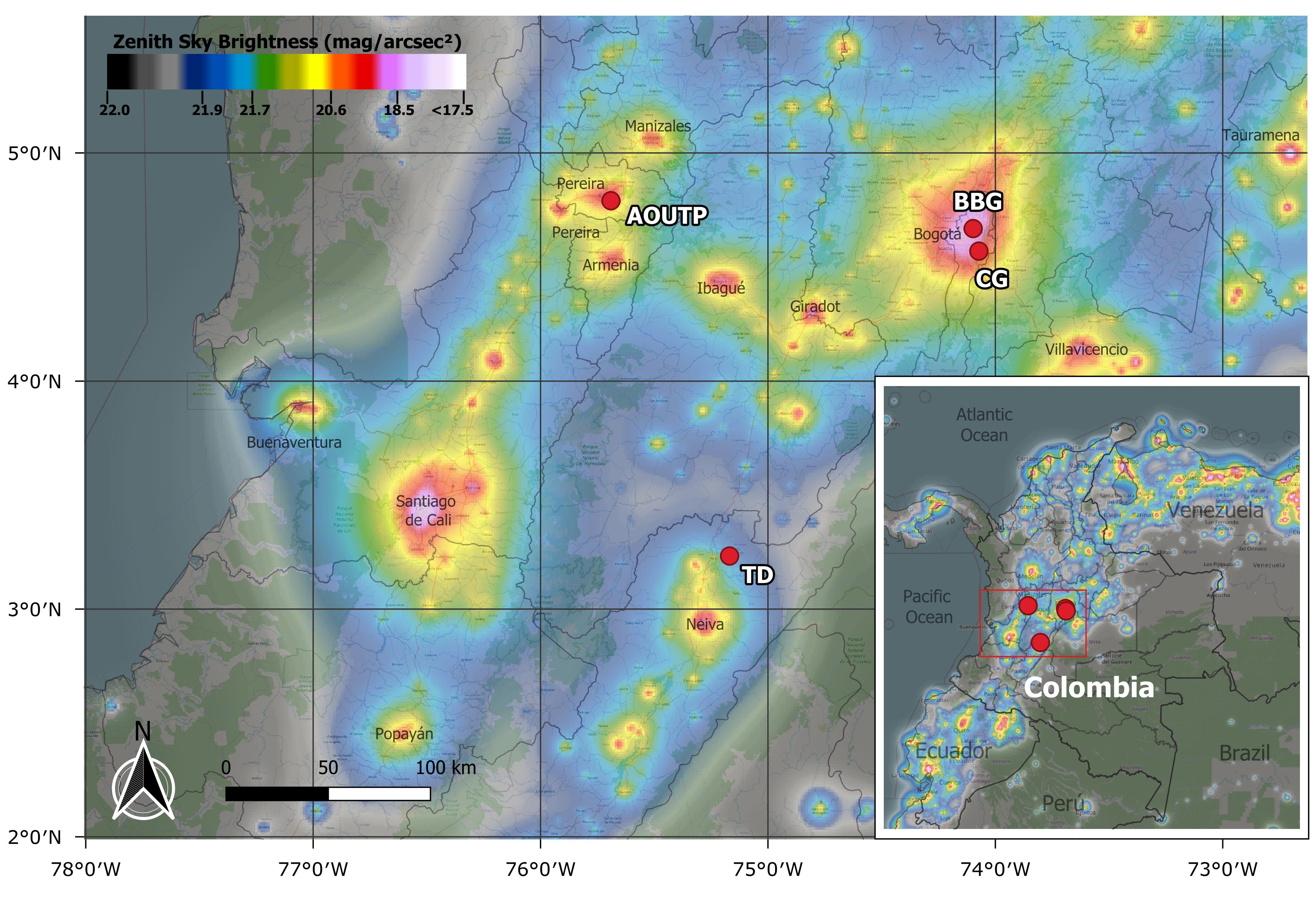}
    \caption{The geographical distribution of the four sites analyzed in this study is represented by red points with white labels on a background color map, constructed using data from the New World Atlas of Artificial Night Sky Brightness \citep{falchi2016new}.}
    \label{fig:map}
\end{figure*}

Several sites were selected to measure NSB in Colombia based on previous SQM records. These locations include the BBG and the TD, both of which have existing SQM data (\citealt{urrego2016analisis} and \citealt{goez2021comparative}, respectively). The study also incorporates the Astronomical Observatory of the Universidad Tecnol\'ogica de Pereira (hereafter AOUTP) and Cerro Guadalupe (hereafter CG) in Bogot\'a. Each site has distinct characteristics in terms of topography and primary activities. AOUTP is focused on astronomical research, TD is known for astrotourism, BBG is dedicated to biodiversity conservation, and CG serves as an ecological corridor\footnote{An ecological corridor is defined by the Central American Commission on Environment and Development as \textquotedblleft a delimited geographical space that provides connectivity between landscapes, ecosystems, and habitats, whether natural or modified, and ensures the maintenance of biological diversity as well as ecological and evolutionary processes\textquotedblright.}. Figure \ref{fig:map} depicts the geographic locations of these sites overlaid on a pseudo-color map adapted from the New World Atlas of Artificial Night Sky Brightness \citep{falchi2016new}. Table \ref{Tab:Coor-Geo} presents the geographic coordinates obtained and the acronyms by which each site is identified.

\begin{table*}[!t]
\centering
  \newcommand{\DS}{\hspace{6\tabcolsep}} 
  \setlength{\tabnotewidth}{0.5\textwidth}
  \setlength{\tabcolsep}{0.4\tabcolsep}
  \tablecols{6}
  \caption{Geographic coordinates} 
  \label{Tab:Coor-Geo} 
  \begin{tabular}{c c c c c c}
    \toprule
    \multicolumn{1}{c}
    {\textbf{Site}} & \multicolumn{1}{c}
    {\textbf{Acronym}} & \multicolumn{1}{c}
    {\textbf{City}} & \multicolumn{1}{c}{\begin{tabular}[c]{@{}c@{}}\textbf{Latitude~} \\
    \textbf{[$\degree \; \arcmin \; \arcsec$] N}\end{tabular}} &\multicolumn{1}{c}{\begin{tabular}[c]{@{}c@{}}\textbf{Longitude~}\\
    \textbf{[$\degree \; \arcmin \; \arcsec$] W}\end{tabular}} & \multicolumn{1}{c}{\begin{tabular}[c]{@{}c@{}}\textbf{Elevation~}\\
    \textbf{[masl]}\end{tabular}}\\
    \midrule
    \multicolumn{1}{c}{\begin{tabular}[c]{@{}c@{}}Astronomical ~\\Observatory UTP\end{tabular}}  & AOUTP & Pereira  & 04 47 26.06 & 75 41 25.57 & 1514 \\
    \cmidrule{1-6}
    \multicolumn{1}{c}{\begin{tabular}[c]{@{}c@{}}Tatacoa ~\\Desert \end{tabular}}  & TD & Villavieja & 03 13 51.98 & 75 09 16.22  & 481 \\
    \cmidrule{1-6}
    \multicolumn{1}{c}{\begin{tabular}[c]{@{}c@{}}Bogot\'a ~\\Botanical Garden \end{tabular}} & BBG & Bogot\'a & 04 40 03.49 & 74 06 02.16 & 2565 \\
    \cmidrule{1-6}
    \multicolumn{1}{c}{\begin{tabular}[c]{@{}c@{}}Cerro ~\\Guadalupe \end{tabular}} & CG & Bogot\'a & 04 35 29.20 &  74 03 12.86 & 3273 \\
    \bottomrule  
  \end{tabular}
\end{table*}

\subsection{Data Acquisition, reduction and analysis} \label{Sec_2.2}
The SQC used in this study is a Canon EOS 6D Mark II featuring a 26.2-megapixel full-frame CMOS sensor and an integrated GPS. It comes equipped with a Sigma 8mm EX DG fisheye lens able to image in a single shot a field of view of 186°. The entire setup is mounted on a Manfrotto MK055XPRO3-3W tripod. After several years of intensive use, we can confirm that the SQC maintains a photometric precision of $\pm 0.01$ mag/arcsec$^2$, a CCT precision of $\pm$ 30\ K, and a spatial resolution of $\pm$ 0.02°.

We took data only after astronomical twilight, i.e., when the Sun elevation was $\geq -$18\degree, and on those times when the Moon was $\geq$10\degree \ below the horizon. At each site, we identified the optimal location for the SQC installation, ensuring a 30 meter radius free of nearby luminaires and avoiding any direct illumination on the equipment. Additionally, we prioritized sites with the clearest possible view of the sky to minimize (whenever possible) obstructions from elements such as trees, mountains, and buildings. As a matter of fact, due to the natural surroundings, achieving a fully unobstructed view was not always feasible, particularly at BBG and CG. 
Subsequently, the camera was positioned to point towards the zenith, with its front aligned to geographical north using a compass. A circular bubble level was placed on the cover to ensure precise leveling. In this study, the ISO was set to 1600 at all sites, while the shutter speed was adjusted based on ambient brightness to prevent saturation and ensure a signal to noise ratio (S/N) greater than 20. All this information was recorded in the Observation Log, shown here as Table \ref{Tab:JO}, which includes acronyms, date, time, exposure duration, and S/N values. 

In order to preserve the maximum amount of information, the images were processed and analyzed in the native raw Canon .CR2 format using the commercial SQC software (v.1.9.9). After each image we also took the corresponding dark, that is automatically subtracted before the image file is written to disk. For further details about the data acquisition, reduction and analysis processes, the interested reader can refer to \citet{angeloni2024toward}.

\begin{table*}[!t]
\centering
  \newcommand{\DS}{\hspace{6\tabcolsep}} 
  \setlength{\tabnotewidth}{0.68\textwidth}
  \setlength{\tabcolsep}{0.9\tabcolsep}
  \tablecols{6}
  \caption{Observation Log} 
  \label{Tab:JO}
  \begin{tabular}{c c c c c c}
    \toprule
    \multicolumn{1}{c}{\begin{tabular}[c]{@{}c@{}}\textbf{Site~}\\\textit{acronym}\end{tabular}}
    & \multicolumn{1}{c}{\begin{tabular}[c]{@{}c@{}}\textbf{Date~}\\\textit{aaaa-mm-dd}\end{tabular}} & \multicolumn{1}{c}{\begin{tabular}[c]{@{}c@{}}\textbf{Time~}\\\textit{local}\end{tabular}} 
    & \textbf{ISO} 
    & \multicolumn{1}{c}{\begin{tabular}[c]{@{}c@{}}\textbf{Exp. Time~}\\
    \textit{[s]}\end{tabular}} 
    & \textbf{S/N}\\
    \midrule
    AOUTP & 2022-09-15 & 19:24:07 & 1600 & 5 &  35 \\
    \cmidrule{1-6}
    TD & 2022-09-18 & 22:07:27 & 1600 & 150 & 22 \\
    \cmidrule{1-6}
    BBG & 2022-09-20 & 19:18:22 & 1600 & 3 & 41 \\
    \cmidrule{1-6}
    CG  & 2022-09-20 & 22:45:29 & 1600 &  11 & 33\\
    \bottomrule
  \end{tabular}
\end{table*}

\section{Results}\label{Sec_3}

This section provides a comprehensive description of each measurement location, emphasizing its unique characteristics and presenting the results of the observations. Considering that Colombia is one of the cloudiest regions in the world \citep{poveda2011hydro}, given its average annual cloud cover of 68.6\% \footnote{https://www.extremeweatherwatch.com/countries/colombia}, it is essential to consider clouds as an integral part of the natural landscape when analyzing our SQC data. 

Therefore, for each site, an SQC image is presented, consisting of four components (for example, Figure \ref{fig:sqc_OAUTP}): the topocentric RGB map, the Cloud Map, the NSB map, and the CCT map. These images are arranged as follows: the RGB map is in the top left, the Cloud Map is in the bottom left, the NSB map is in the top right, and the CCT map is in the bottom right. In the RGB maps, the colored spherical triangles encapsulate the regions of interests, while in the remaining maps, the orange (yellow) line represents the galactic (ecliptic) plane, while the purple ring corresponds to the 30° elevation ring, which serves as a photometric indicator described in more detail in Section \ref{Sec_4.1}. The cloud map employs a color scale to depict sky conditions: blue indicates clear skies, red signifies cloudiness, and green denotes the horizon. The NSB map includes the coordinates of the darkest point in the sky, specified in terms of Azimuth (AZ) and Zenith Angle (ZA), along with its corresponding NSB value. The upper scale bar displays NSB values in $[V \ mag/arcsec^2]$, while the lower scale bar shows CCT values in [K]. Under very bright skies the Las Vegas scale (NSB values ranging between 11 and 21 $[V \ \mathrm{mag/arcsec^2}]$) is displayed instead of the usual one (that goes between 14 and 24 $[V \ \mathrm{mag/arcsec^2}]$).

Subsequently, the azimuth profiles (see Figures \ref{fig:UTP_Horizontal}, \ref{fig:DT_Horizontal}, \ref{fig:JBB_Horizontal}, and \ref{fig:CG_Horizontal}) are presented, constructed using a 1°-wide ring. The altitude values are selected based on the horizon of each image. In the graphs, the NSB profile is displayed on the left and the CCT profile on the right, with the names of the main peaks in the NSB and CCT values labeled for reference.

On the other hand, the zenith profiles (see Figures \ref{fig:UTP_Vertical}, \ref{fig:DT_Vertical}, \ref{fig:JBB_Vertical}, and \ref{fig:CG_Vertical}) are constructed considering all zenith angles. However, the amplitude of these profiles varies according to the azimuthal range specified in Table \ref{Tab:Areas}, as each region of interest covers a different extent depending on its influence on the site. The arrangement of the zenith profiles follows the same format as the azimuth profiles, with the NSB profile on the left and the CCT profile on the right.

Table \ref{Tab:Areas} details the azimuthal ranges of interest and the most significant sources of ALAN for each site, along with their corresponding distances. To identify these sources, we utilized Colombia's administrative divisions, starting with visible cities, followed by communes\footnote{The National Administrative Department of Statistics (DANE, for its acronym in Spanish) defines a commune as an administrative subdivision of a city or municipality that groups several neighborhoods to facilitate local management, community participation, and the administration of public services.} and rural districts (vereda). In the case of Bogot\'a, as a capital district, the administrative divisions refer to localities.
The \textquotedblleft Region\textquotedblright \ column in the table employs a color code to represent the brightest ALAN sources at each site: blue indicates the brightest source, followed by red, green, and black. For the TD site, orange represents an ALAN-free
direction, defined as an azimuthal range where no ALAN sources are detected within a distance of 150 km \citep{angeloni2024toward}.
\begin{table*}
\centering
\newcommand{\DS}{\hspace{5\tabcolsep}} 
  \setlength{\tabnotewidth}{\textwidth}
  \setlength{\tabcolsep}{0.5\tabcolsep}
  \tablecols{5}
  \caption{ALAN Sources\tabnotemark{a}}   
  \label{Tab:Areas} 
\resizebox{\textwidth}{!}{%
  \begin{tabular}{c c c c c }
    \toprule
    \multicolumn{1}{c}{\textbf{Site}} & \multicolumn{1}{c}{\textbf{Region}} & \multicolumn{1}{c}{\begin{tabular}[c]{@{}c@{}}\textbf{Azimuth Range~} \\
    \textbf{[$\degree$]}\end{tabular}} & \multicolumn{1}{c}{\begin{tabular}[c]{@{}c@{}}\textbf{ALAN~} \\
    \textbf{\textit{Main Sources}}\end{tabular}}  & \multicolumn{1}{c}{\begin{tabular}[c]{@{}c@{}}\textbf{Distance~}\\ \textbf{[\textit{km}]}\end{tabular}}\\
    \midrule  
    \multirow{10}{*}{AOUTP} & \multirow{3}{*}{\textcolor{blue}{I}} & \multirow{3}{*}{0 - 90} & Universidad Commune  & 1.1 \\
     &  & & Villasantana Commune  & 2.2 \\
     &  & & Dosquebradas  & 3.7 \\
    \cmidrule{2-5}
     & \multirow{3}{*}{\textcolor{red}{II}} & \multirow{3}{*}{270 - 0} & Centro Commune & 2.7 \\ 
     &  &  & Cuba Commune & 5.4 \\  
     &  &  & Mateca\~na Airport & 5.9 \\     
    \cmidrule{2-5}
       & \multirow{2}{*}{\textcolor{green}{III}} & \multirow{2}{*}{180 - 270} & Condina Bypass & 3.0 \\
       &  &  & Altagracia Neighborhood & 6.2 \\  
    \cmidrule{2-5}  
      & \multirow{2}{*}{\textcolor{black}{IV}} & \multirow{2}{*}{90 - 180} & Vereda Mundo Nuevo & 1.3 \\
       &  &  & El Remanso Neighborhood  & 2.1 \\     
    \cmidrule{1-5}  
     \multirow{8}{*}{TD} & \multirow{2}{*}{\textcolor{blue}{I}} & \multirow{2}{*}{178 - 238} & Ecopetrol Dina Field  & 23.4  \\
       &  &  & Neiva  & 36.8 \\
    \cmidrule{2-5}
    & \multirow{3}{*}{\textcolor{red}{II}} & \multirow{3}{*}{249 - 290} & Villavieja & 7.2\\
     & & & Aipe & 9.5 \\
     & & & 3C LTDA (Agro-industrial company)   & 11.5 \\
    \cmidrule{2-5}
     & \multirow{3}{*}{\textcolor{green}{III}} & \multirow{4}{*}{11 - 45} & Melgar  & 121 \\
     & & & Girardot  & 125 \\
     & & & Bogot\'a  & 193 \\
    \cmidrule{2-5}
       & \multirow{2}{*}{\textcolor{orange}{IV}} & \multirow{2}{*}{140 - 145} & ALAN-free direction \multirow{2}{*}{} & \multirow{2}{*}{\nodata} \\
       &  &  & (Cordillera de los Picachos National Park)    &  \\  
    \cmidrule{1-5}  
     \multirow{8}{*}{BBG} & \multirow{2}{*}{\textcolor{blue}{I}} & \multirow{2}{*}{180 - 270} & Kennedy Locality & 6.7 \\
      &  &  & Ciudad Bolivar Locality & 13 \\
     \cmidrule{2-5}
      & \multirow{2}{*}{\textcolor{red}{II}} & \multirow{2}{*}{90 - 180} & Teusaquillo Locality & 3.4 \\
      &  &  & Chapinero Locality & 13  \\
      \cmidrule{2-5}
       & \multirow{2}{*}{\textcolor{green}{III}} & \multirow{2}{*}{270- 360} &  Fontib\'on Locality & 3.5 \\
      &  &  & Engativ\'a Locality  & 5.0 \\
      \cmidrule{2-5}
      & \multirow{2}{*}{\textcolor{black}{IV}} & \multirow{2}{*}{0 - 90} &
      Suba Locality & 7.2 \\
      &  & & Usaqu\'en Locality & 8.0 \\   
    \cmidrule{1-5}
    \multirow{8}{*}{CG} & \multirow{2}{*}{\textcolor{blue}{I}} & \multirow{2}{*}{180 - 270} & Usme Locality  & 10 \\
     &  & & Ciudad Bolivar Locality  & 11.5 \\
    \cmidrule{2-5}
      & \textcolor{red}{II} & 270 - 0 & Bogot\'a  & 3 \\
    \cmidrule{2-5}
     & \multirow{2}{*}{\textcolor{green}{III}} & \multirow{2}{*}{90 - 180} & Choach\'ia & 11 \\
     &  &  & Chingaza National Park& 34 \\ 
    \cmidrule{2-5}
     & \multirow{2}{*}{\textcolor{black}{IV}} & \multirow{2}{*}{0-90} & Usaqu\'en Locality & 13.0 \\
     &  & & La Calera  & 16.7 \\
    \bottomrule
    \tabnotetext{a}{ALAN sources for each site: The numbers I, II, III, and IV correspond to the study regions, with colors coding the region with the highest contribution of artificial brightness. All sources are arranged in ascending order concerning the measurement site. For detailed information, please refer to the main text.} 
  \end{tabular}
}
\end{table*}

\subsection{Astronomical Observatory of the UTP (AOUTP)} \label{Sec_OAUTP}

The AOUTP\footnote{https://observatorioastronomico.utp.edu.co/} is located in the Central West metropolitan area\footnote{The fifth largest in Colombia, with a population of 735,796 according to the 2018 census by DANE.}. Specifically, the observatory is situated in the southeastern part of Pereira, on the university campus in the La Julita sector. The observatory has received several international recognitions. Notably, it has been granted the code W63 by the Minor Planet Center, identifying it as an observatory authorized to obtain astrometric data of minor bodies in the Solar System \citep{villariaga2017obtaining}. Additionally, the observatory holds the UTP 0383 code from the Stanford Solar Center for space weather monitoring \citep{galvis2019space}. These recognitions establish the AOUTP as one of Colombia's leading scientific centers for astronomical and space science.

In 2025, the observatory will expand its infrastructure and upgrade its astronomical equipment. A new facility will house a 70 cm diameter telescope, which, along with the existing instrumentation, will establish the observatory as the leading center for astronomical research in Colombia. These advancements will not only elevate the university's academic contributions but also strengthen its research capabilities, community outreach, and engagement in the field of astronomy.

It is worth noticing the ecological value of the UTP campus, where 58\% of its surface is dedicated to forest conservation and consequently classified as Natural Wild by Botanic Gardens Conservation International\footnote{https://www.bgci.org/}, as plant species thrive in their natural environment without human intervention. Specifically, 13 hectares are allocated to the UTP Botanical Garden, which houses approximately 542 species of flora that are meticulously identified and protected \citep{garcia2011estudio}. This area serves as a valuable resource for research and environmental education and a space for recreation and engagement with nature for students and visitors. The conservation and sustainability initiatives implemented on campus have garnered national and international recognition, establishing UTP as a model in university environmental management.

\begin{figure*}
\centering
    \includegraphics[width=\textwidth]{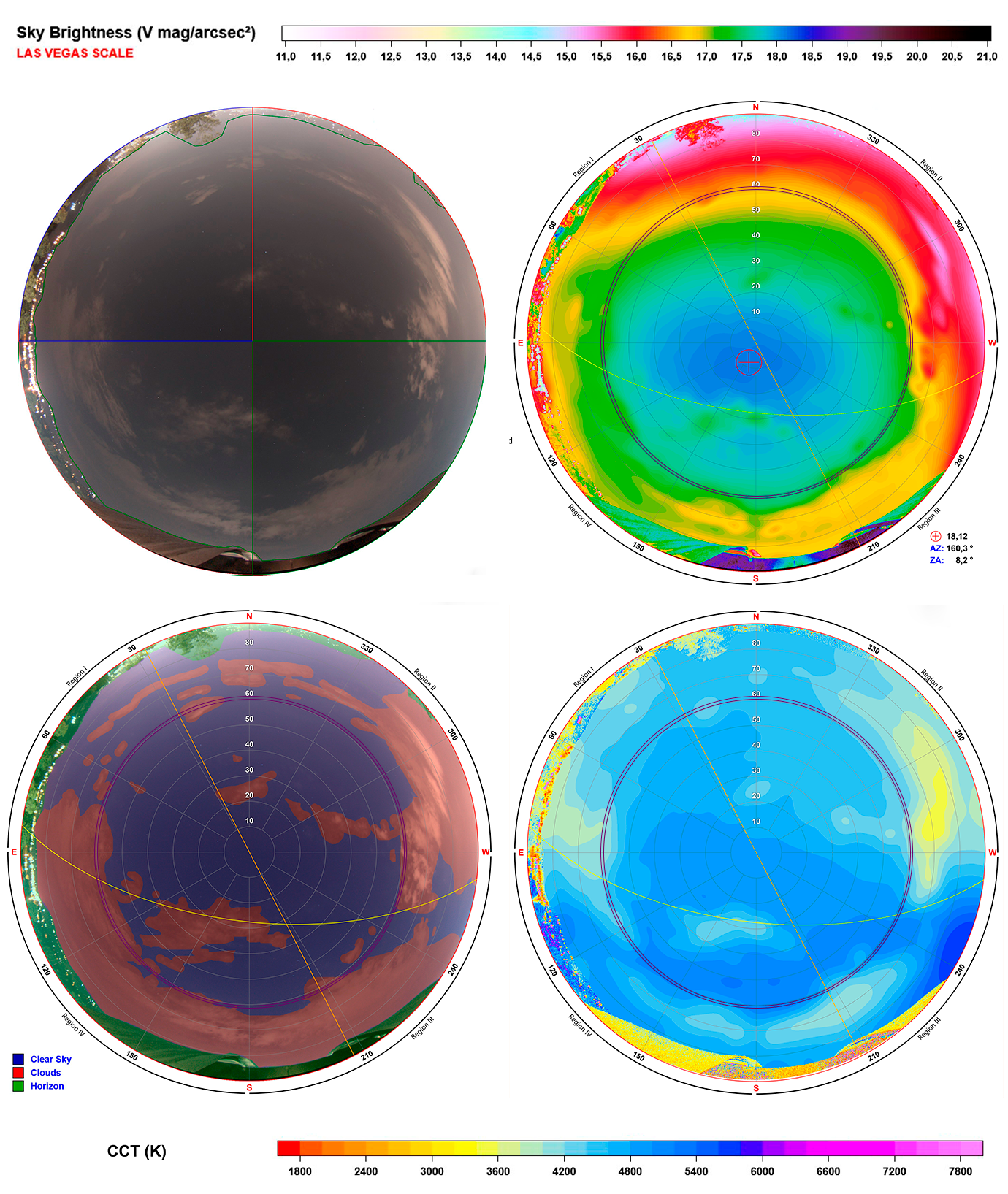}
       \caption{SQC Images of the Night Sky at AOUTP captured on September 15, 2022. North is to the top and East is to the left. The upper-left panel displays the RGB image, with spherical triangles of color-coded azimuth ranges as presented in Table \ref{Tab:Areas} and discussed in Section \ref{Sec_OAUTP}. The upper-right panel shows the NSB map in V $mag/arcsec^2$ and includes the coordinates of the darkest point along with its corresponding NSB value ($\oplus$). The lower-left panel presents the cloud map, highlighting cloudy regions in red. The lower-right panel shows the CCT map in Kelvin. In the last three panels, the orange (yellow) line corresponds to the galactic (ecliptic) plane, while the purple ring corresponds to the 30° elevation ring (refer to Section \ref{Sec_4.1}). Under very bright skies the Las Vegas scale (NSB values ranging between 11 and 21 $[V \ \mathrm{mag/arcsec^2}]$) is displayed instelad of the usual one (that goes between 14 and 24 $[V \ \mathrm{mag/arcsec^2}]$).}
    \label{fig:sqc_OAUTP}
\end{figure*}

Figure \ref{fig:sqc_OAUTP} presents an all-sky view from the AOUTP observatory. In the RGB map, the observatory's dome is visible on the horizon to the south, while the eastern horizon reveals the surrounding mountains and the lights from nearby houses. The sky features several cloudy regions illuminated by the skyglow from the city of Pereira, as corroborated by the cloud map. The NSB map highlights a bright region extending approximately from $240^{\circ}$ to $60^{\circ}$ in azimuth, corresponding to the cities of Pereira and Dosquebradas. Furthermore, a zone of increased surface brightness is detected in the azimuthal range of $270^{\circ}$ to $300^{\circ}$, exhibiting a lower CCT than the rest of the sky in the CCT map, indicating the presence of a cloud amplifying the city's surface brightness. The combination of these images provides valuable insights into how atmospheric conditions and urban lighting influence sky quality and, consequently, the effectiveness of astronomical observations.

\begin{figure*}[!t]
  \includegraphics[width=0.45\textwidth]{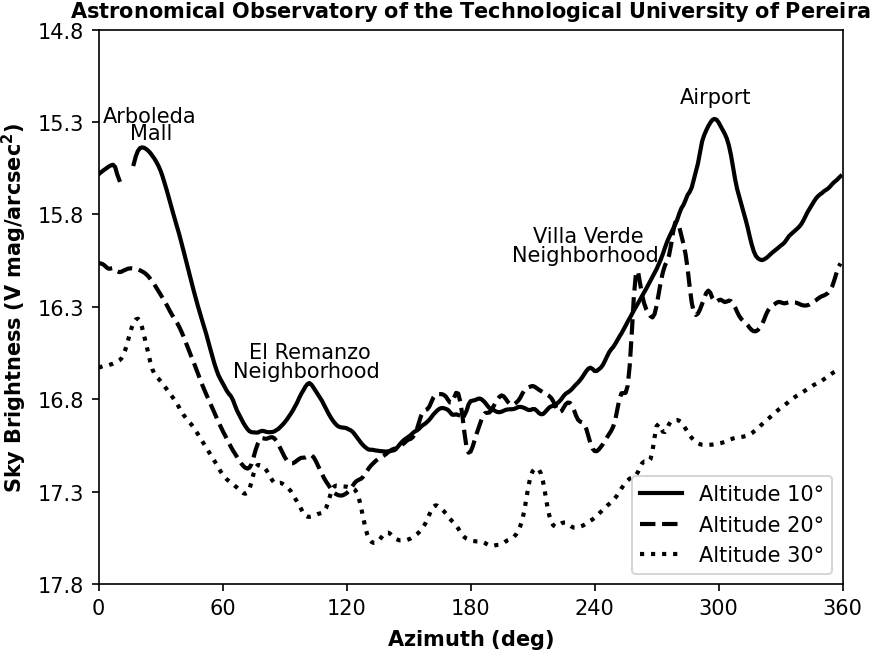}
  \hspace*{\columnsep}%
  \includegraphics[width=0.45\textwidth]{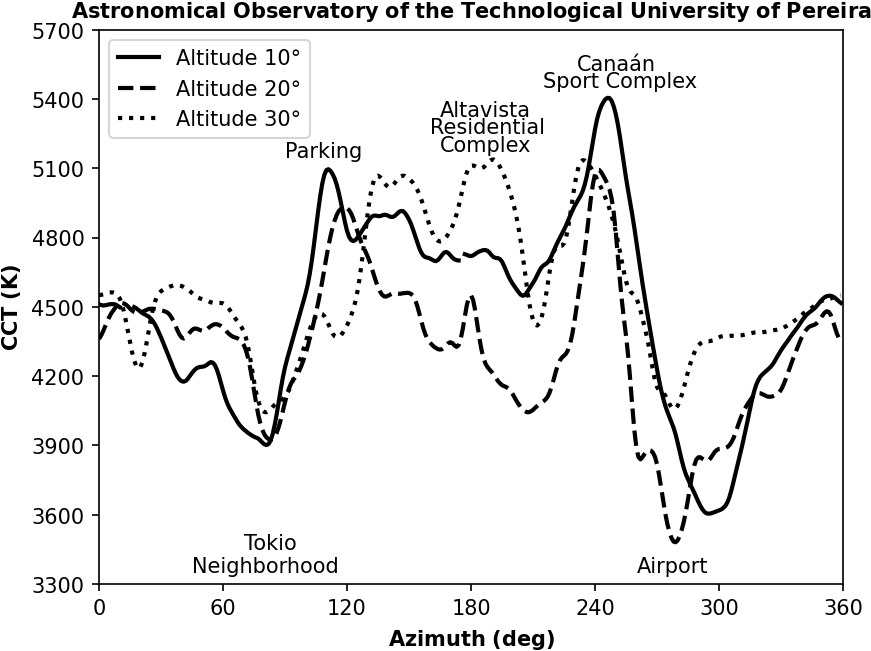}
  \caption{Azimuthal NSB/CCT profiles (left/right panel) averaged within $1^{\circ}$ wide rings centered at $10^{\circ}$, $20^{\circ}$, and $30^{\circ}$ elevation at AOUTP. These profiles identify the local maxima of NSB and CCT, indicating the azimuthal directions of the ALAN sources. For more details, refer to Figure \ref{fig:sqc_OAUTP} and Table \ref{Tab:Areas}.}
  \label{fig:UTP_Horizontal}
\end{figure*}

To identify the areas or sites that contribute most to the NSB and CCT, we constructed rings at different elevations to determine the directions of the most significant contributing sources in each study region. We refer to these segments of the sky as azimuthal profiles, abbreviated as AZ. Figure \ref{fig:UTP_Horizontal} presents the azimuthal profile for the AOUTP. Using the ALAN sources from Table \ref{Tab:Areas}, we determined the azimuth of the highest contribution to NSB and CCT for each region:

\begin{itemize}
    \item Region I exhibit a NSB peak at  $AZ\approx21^{\circ}$, corresponding to the direction of the Arboleda Mall, and a CCT minimum at $AZ\approx83^{\circ}$, corresponding to the Tokio neighborhood.
    \item Region II shows a maximum NSB at $AZ\approx298^{\circ}$  and a minimum CCT at $AZ\approx280^{\circ}$, corresponding to the airport direction.
    \item Region III displays a maximum NSB at $AZ\approx261^{\circ}$, corresponding to the Villa Verde neighborhood and two CCT maxima at $AZ\approx192^{\circ}$ and $AZ\approx242^{\circ}$, corresponding to the Altavista residential complex and the Canaán Sports Complex.
    \item Region IV has a maximum NSB at $AZ\approx102^{\circ}$, corresponding to the sports complex of the El Remanso neighborhood, and a maximum CCT at $AZ\approx119^{\circ}$, corresponding to the parking of the Botanical Garden.
\end{itemize}

 The primary contributors to the NSB for the AOUTP are the Arboleda Mall and the airport, which typically operate almost until midnight. The airport's lighting remains constant to ensure safe landings, while the mall maintains its lights for various events. In contrast, the maximum CCT values correspond to areas near the AOUTP, such as the parking of the Botanical Garden, sports complex, and residential complexes. These places typically use LED lights and, in many cases, have lighting issues such as over-illumination or improper light orientation, which disperses light both toward the ground and the sky.

\begin{figure*}[!t]
  \includegraphics[width=0.45\textwidth]{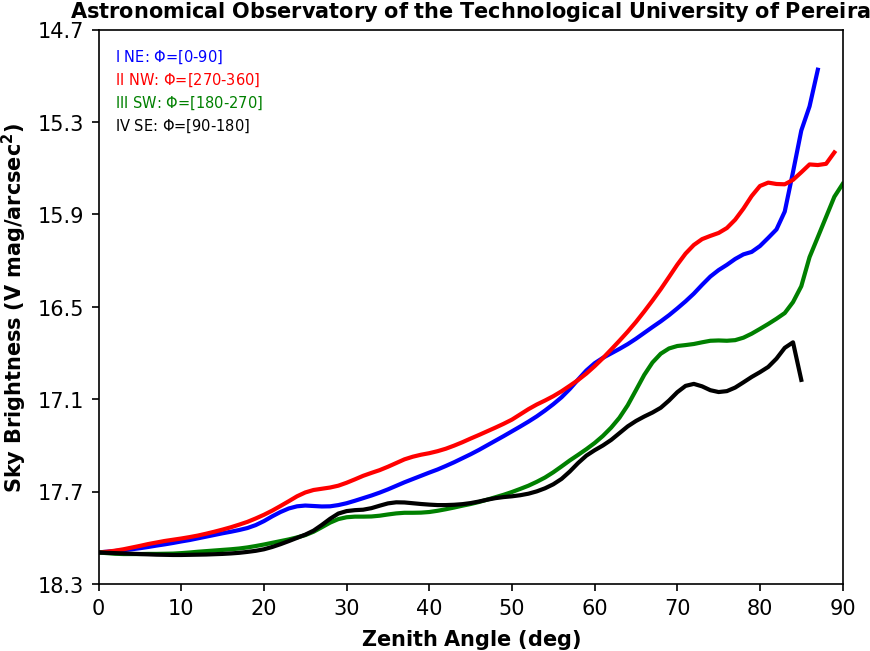}%
  \hspace*{\columnsep}%
  \includegraphics[width=0.45\textwidth]{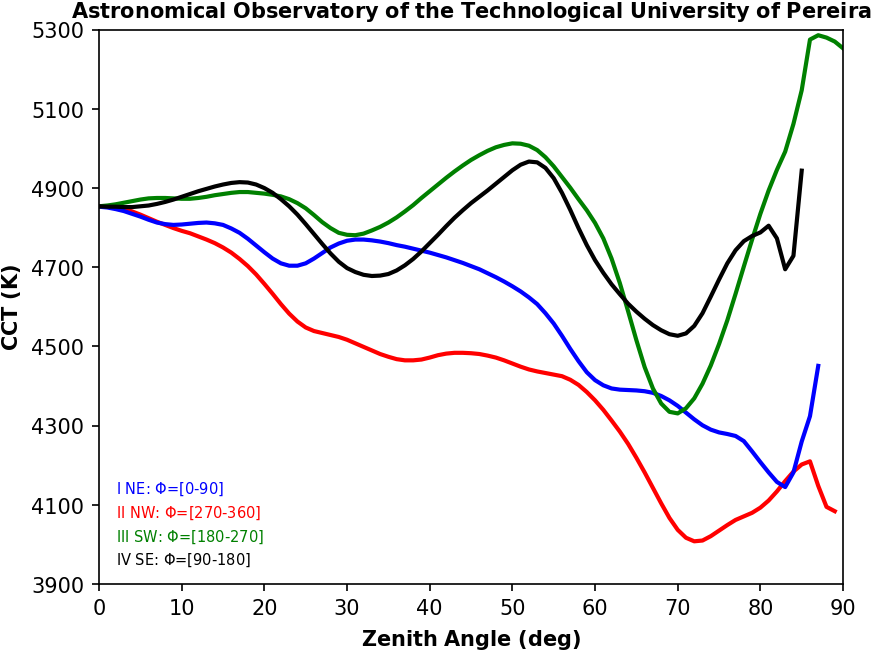}
  \caption{Zenith profiles, NSB/CCT variance as a function of zenith angle (left panel/right panel). For details, refer to Section \ref{Sec_OAUTP}, Figure \ref{fig:sqc_OAUTP}, and Table \ref{Tab:Areas}.}
  \label{fig:UTP_Vertical}
\end{figure*}

To determine the altitude affected by the AOUTP, we constructed a ZA profile (see Figure \ref{fig:UTP_Vertical}), which revealed distinct brightness patterns across different regions as they approach the horizon. In Region I, a sharp increase in brightness is observed at ZA = 83°, attributed to the city glow. Region II exhibits a rise in brightness between ZA = 73° and ZA = 81°, likely caused by a cloud amplifying the NSB. In Region III, a slight increase is evident  between ZA = 25° and ZA = 40°, associated with the presence of a cloud. Additionally, a secondary increase occurs at ZA = 83°, linked to urban glow. In Region IV, NSB increases between ZA = 26° and ZA = 43°, due to cloud covered, and again between ZA = 75° and ZA = 85°, potentially caused by urban glow. It is worth noting that some areas do not extend to the horizon, as they are physically obstructed by features like mountains or the dome.

\subsection{Tatacoa Desert (TD)} \label{Sec_DT}

The Tatacoa Desert is a tropical dry forest situated in the Magdalena River valley, within the department of Huila, approximately 10 km from the municipality of Villavieja\footnote{In 2018, the DANE census recorded a population of 7,308 inhabitants.}. It is the second driest region in Colombia, renowned for its eroded landscape, forming a labyrinth of canyons, valleys, and rock formations in striking red and gray hues. The Tatacoa Desert uniquely combines geological, climatic, and biogeographical factors, making it an ideal site for studying Neogene biodiversity in South America (\citealt{florez2013paleosuelos}, \citealt{montes2021middle}) and paleogeography \citep{dill2020badland}.

As one of Colombia's top tourist destinations, the TD attracts visitors with its semi-arid landscapes, outdoor activities, including stargazing \citep{vega2022retos}. In September 2019, it was certified as a Starlight Destination\footnote{https://fundacionstarlight.org/}, the first site in Colombia to receive this recognition. The Starlight Foundation, with support from UNESCO, the World Tourism Organization (UNWTO), and the International Astronomical Union (IAU), grants this certification to areas that protect and conserve dark skies.

Today, the Tatacoa Desert is a hub for astrotourism and a premier destination in Colombia for astronomy enthusiasts. The desert hosts several observatories and astronomical complexes. For our measurements, we were at the Astrosur Observatory\footnote{https://www.facebook.com/Tatacoa.Astronomia/}, located approximately in the center of the desert, it provided the necessary safety conditions to conduct our research.

TD is one of the locations that presented a NSB measurement with the SQM instrument during the night of April 14-15, 2014. These measurements were made during a total lunar eclipse, and therefore with a full Moon, allowing for a record of approximately 9 hours of measurement. Each measurement was taken at 5-minute intervals; for more details, refer to \citet{goez2021comparative}. As this is the only existing record of sky quality for this location, we will use the darkest measurement, which is 21.26 $mag/arcsec^2$, as the reference value. This reference will be fundamental for evaluating the evolution of sky quality in future studies.

\begin{figure*}
    \centering
    \includegraphics[width=\textwidth]{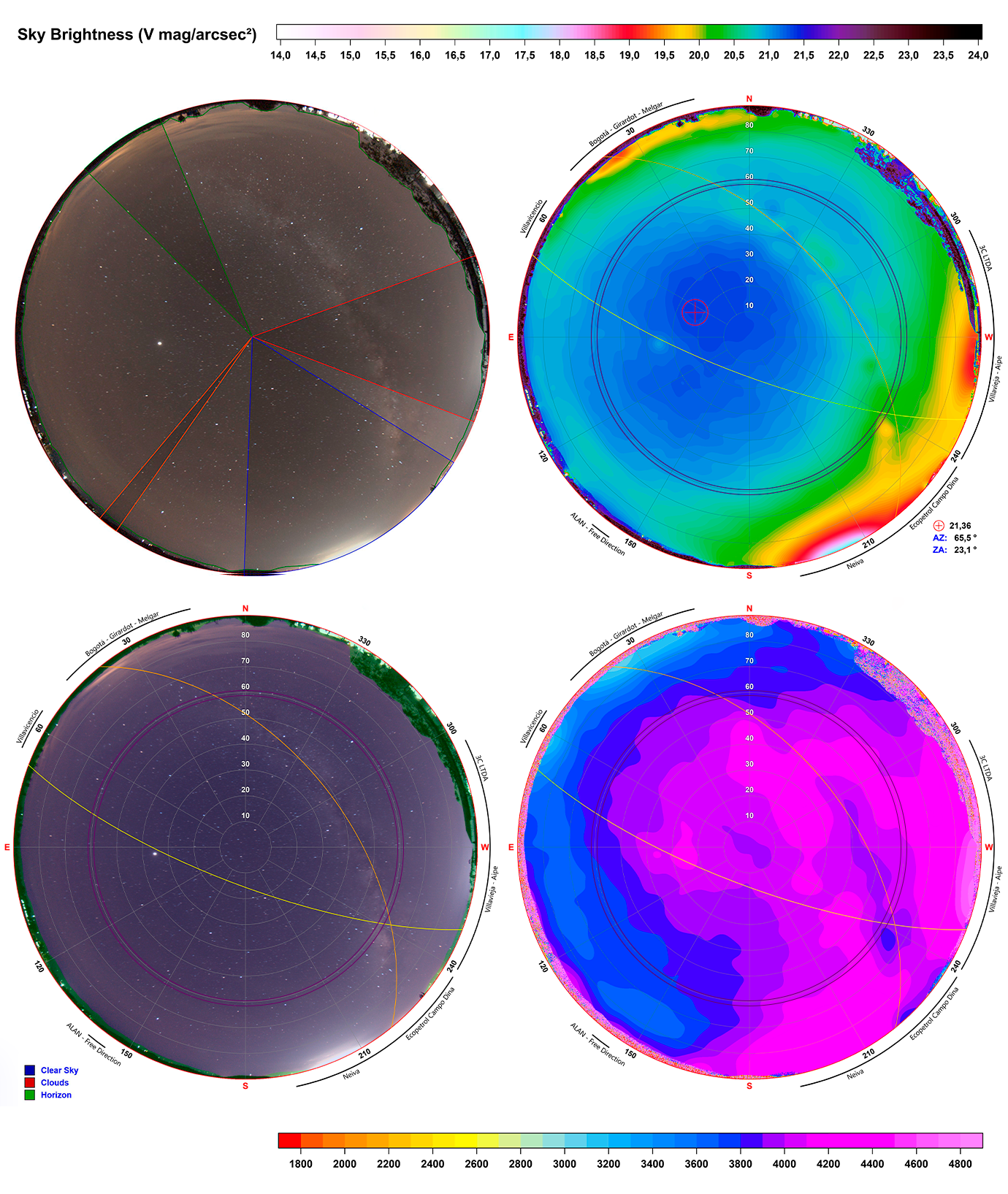}
    \caption{Same as Figure \ref{fig:sqc_OAUTP}, but for TD on September 18, 2022,using the standard scale bar for NSB and a CCT range of [1800–4800 K]. Additionally, the orange spherical triangle corresponds to the ALAN - free direction. For more details, refer to Section \ref{Sec_DT} and Table \ref{Tab:Areas}.}
    \label{fig:sqc_DT}
\end{figure*}

The all-sky view from TD is depicted in Figure \ref{fig:sqc_DT}. The upper left image presents an RGB format, where the Milky Way can be clearly seen arcing from north to south, with Jupiter standing out as the brightest point in the sky. TD is the only location that offers an ALAN-free direction. According to the light pollution map, this direction is in the azimuthal range of $140^{\circ}$ to $145^{\circ}$ and corresponds to the Cordillera de los Picachos National Park. This protected area is known for its great diversity of fauna, including endemic and endangered species, and therefore experiences minimal human intervention.

On the horizon, a white glow stands out in the southern and eastern parts of the image, corresponding to the cities of Neiva, located $\sim37$ km away, and Villavieja-Aipe, situated roughly 7 and 10 km, respectively. To the north, another yellow/orange glow originates from the cities of Melgar (121 km), Girardot (125 km), and Bogot\'a (193 km). This glow is intensified by the presence of clouds, which can be identified in the cloud map. The NSB map shows the areas with the highest surface brightness, highlighting that the Milky Way is a natural contribution of light. However, the most intense contributions come from the cities, with Neiva as the main source of brightness. On the other hand, when observing the CCT maps, it can be seen how the CCT of the night sky is confused with the CCT of the cities, with a white glow coming mainly from LED lights.

\begin{figure*}[!t]
  \includegraphics[width=0.45\textwidth]{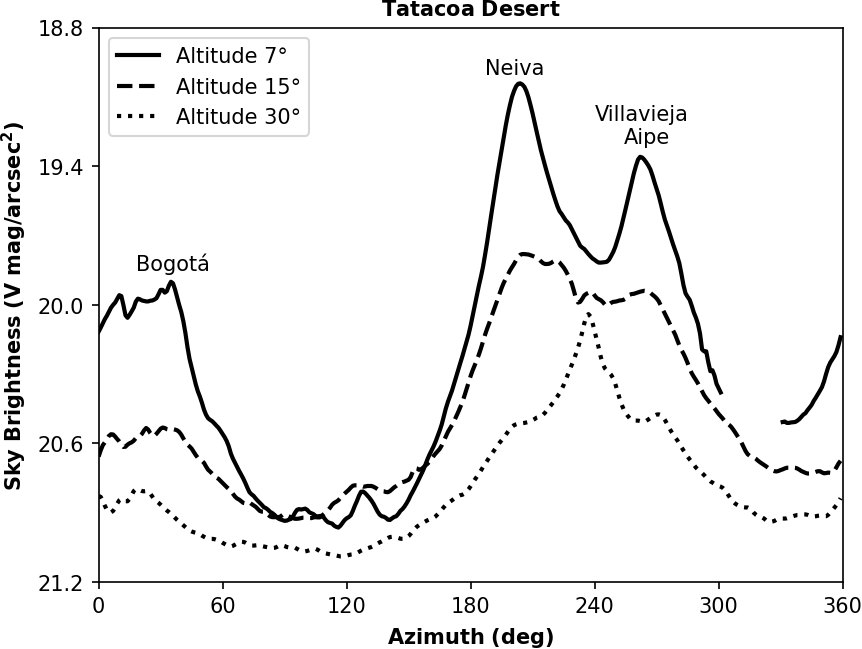}%
  \hspace*{\columnsep}%
  \includegraphics[width=0.45\textwidth]{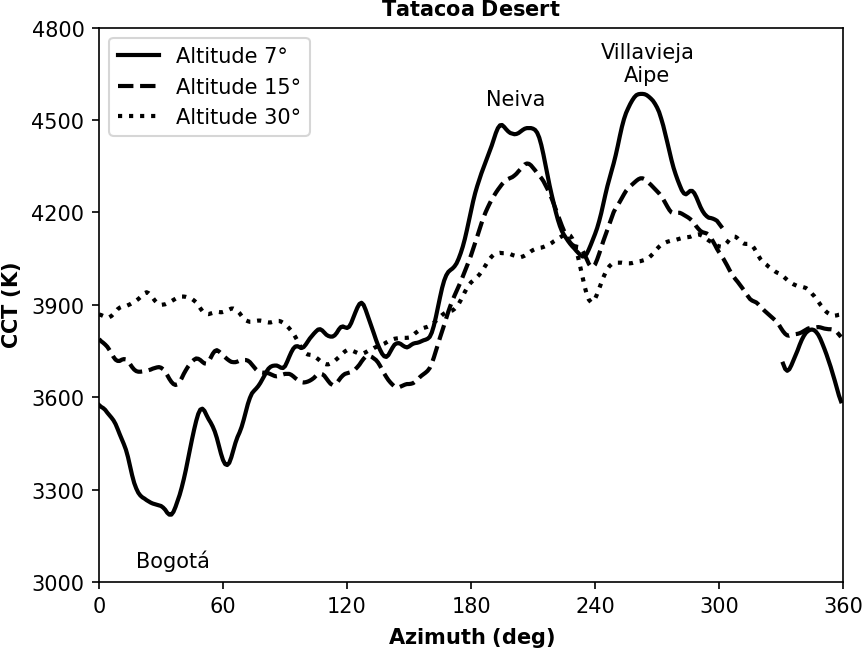}
  \caption{Same as Figure \ref{fig:UTP_Horizontal}, but for TD. In this case, the rings are at $7^{\circ}$, $15^{\circ}$, and $30^{\circ}$ elevation. See Figure \ref{fig:sqc_DT} and Table \ref{Tab:Areas}.}
  \label{fig:DT_Horizontal}
\end{figure*}

Figure \ref{fig:DT_Horizontal} presents the azimuthal profile for TD. In this plot, NSB maxima can be identified for the three regions of interest at $AZ=203^{\circ}$, $AZ=263^{\circ}$, and $AZ=35^{\circ}$, respectively. Similarly, CCT maxima are observed for regions I and II at $AZ=194^{\circ}$ and $AZ=261^{\circ}$, with a minimum for region III at $AZ=34^{\circ}$. The NSB analysis highlights the need to implement controls on the type of illumination in Neiva, Villavieja, and Aipe to prevent the TD sky from losing its dark sky quality or Starlight certification. 

\begin{figure*}[!t]
  \includegraphics[width=0.45\textwidth]{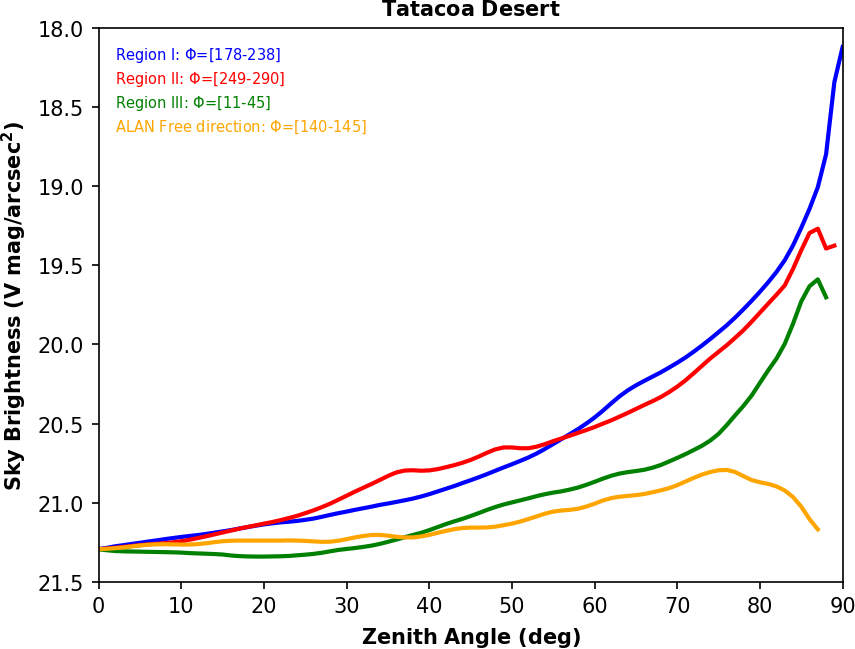}%
  \hspace*{\columnsep}%
  \includegraphics[width=0.45\textwidth]{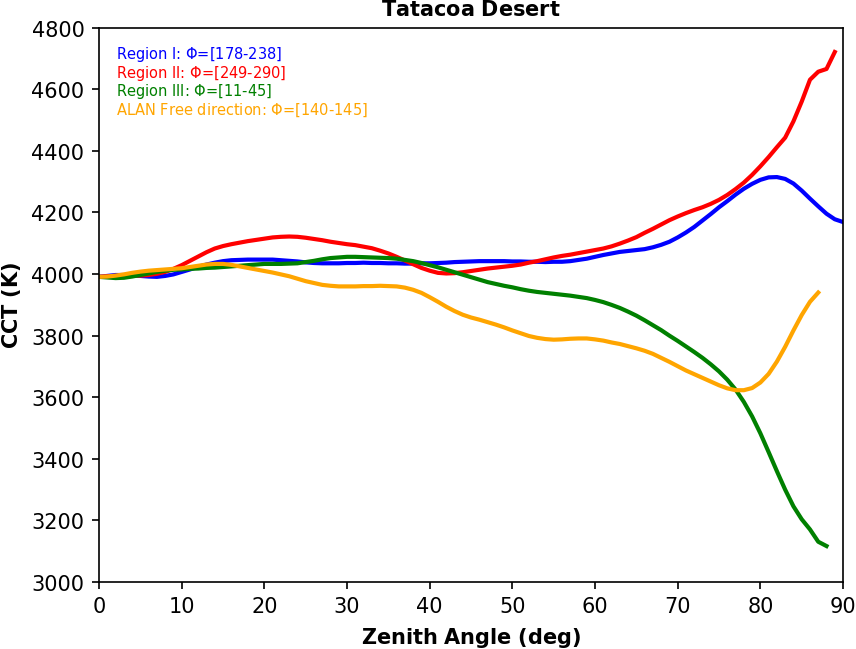}
  \caption{Same as Figure \ref{fig:UTP_Vertical}, but for TD. For more details, see Section \ref{Sec_DT}, Figure \ref{fig:sqc_DT}, and Table \ref{Tab:Areas}.}
  \label{fig:DT_Vertical}
\end{figure*}

Figure \ref{fig:DT_Vertical} presents the zenith profiles of NSB/CCT. In Region I, there is an increase in brightness at  $ZA=64^{\circ}$ due to the passage of the Milky Way, with an additional increase attributed to the city of Neiva. Conversely, in the CCT, no significant variation is observed up to $ZA=69^{\circ}$, also corresponding to Neiva. In Region II, which has the second highest NSB value, an increase is observed as the horizon is approached. Within the range of ZA = 29° to ZA = 53°, the NSB is primarily influenced by the natural contribution of the Milky Way. However, beyond this ZA, the NSB rises rapidly due to the light pollution from the cities of Villavieja and Aipe. Additionally, the CCT analysis indicates that LED lights predominate in this direction, resulting in the highest CCT values observed in this region.

In Region III, the NSB contribution shows a considerable increase at $ZA=71^{\circ}$ due to the ALAN originating from the cities. It is not possible to determine the NSB of the Milky Way in this region since its passage occurs very close to the cities. The CCT shows a noticeable decrease from $ZA=60^{\circ}$, corresponding to the area affected by the city's contributions. Finally, Region IV, the ALAN-free direction is distinguished by its stable NSB profile, identifying it as the darkest area within the studied region. It is important to note, however, that stability in NSB variation does not necessarily indicate low light pollution; an area consistently illuminated by ALAN would also exhibit minimal variation, but with higher brightness levels. The NSB map remains relatively constant up to $ZA=38^{\circ}$. Beyond this angle, an increase in NSB is observed, primarily due to light scattering caused by the city of Neiva. This effect becomes more pronounced at $ZA=48^{\circ}$, where the presence of clouds significantly amplifies the skyglow. Conversely, the CCT profile exhibits a gradual decrease starting around $ZA=14^{\circ}$, reaching its minimum value near $ZA=78^{\circ}$. Beyond this zenith angle, the CCT begins to rise again. This increase is attributed to the absence of clouds over the mountainous region, allowing for clearer visibility of this portion of the sky.

\subsection{Jos\'e Celestino Mutis Bogot\'a Botanical Garden (BBG)} \label{Sec_JBB}

The \textquotedblleft Jos\'e Celestino Mutis\textquotedblright \ Botanical Garden in Bogot\'a is a research center dedicated to conserving high Andean and paramo forest ecosystems. Located in the metropolitan area of Bogot\'a\footnote{In 2018, the DANE census recorded a population of 7.181.469 inhabitants}, in the Engativ\'a locality, the Garden spans approximately 20 hectares. It boasts an impressive collection of flora, featuring 55,000 individual plants representing 1,187 species distributed across 34 ecological zones. The garden also includes various botanical collections, such as ethnobiological, herbarium, carpotheque and tissue collections \citep{vargas2019plantas}.

The BBG has become an environmental reference point and a must-visit destination for those interested in exploring and learning about the city's and country's rich biodiversity. In recent years, it has emerged as one of Bogot\'a's main tourist attractions and serves as a vital green lung in the heart of the metropolis. Additionally, it offers educational programs and interactive activities for visitors of all ages, fostering environmental awareness and respect for nature. The garden also plays a crucial role in scientific research and conservation efforts, collaborating with various institutions to preserve Colombia's unique plant species \citep{cadena2021coleccion}. 

As previously mentioned, BBG had existing ALAN measurements recorded using an SQM. In 2014, a study was conducted using 13 measurement points distributed throughout the garden, starting near the southern part of the lake and following a circular route that ended at the main entrance. For our study, equipped with an SQC instrument, we chose a single measurement point to capture the most comprehensive information of the entire sky. To maximize sky coverage, we strategically selected a location approximately 80 meters from the Tropicarium\footnote{It is a building whose primary function is to protect the flora that grows below 2,400 meters above sea level.} and the greenhouse circuit. This location aligns with an intermediate position between points 4 and 5 from the study by \citet{urrego2016analisis}.

Furthermore, considering the differences in equipment and the SQM's limitation of providing only a single value, we used the average NSB value recorded at the 13 points in the 2014 study during September as a reference. This average value was 16.31 $mag/arcsec^2$.

\begin{figure*}
    \centering
    \includegraphics[width=\textwidth]{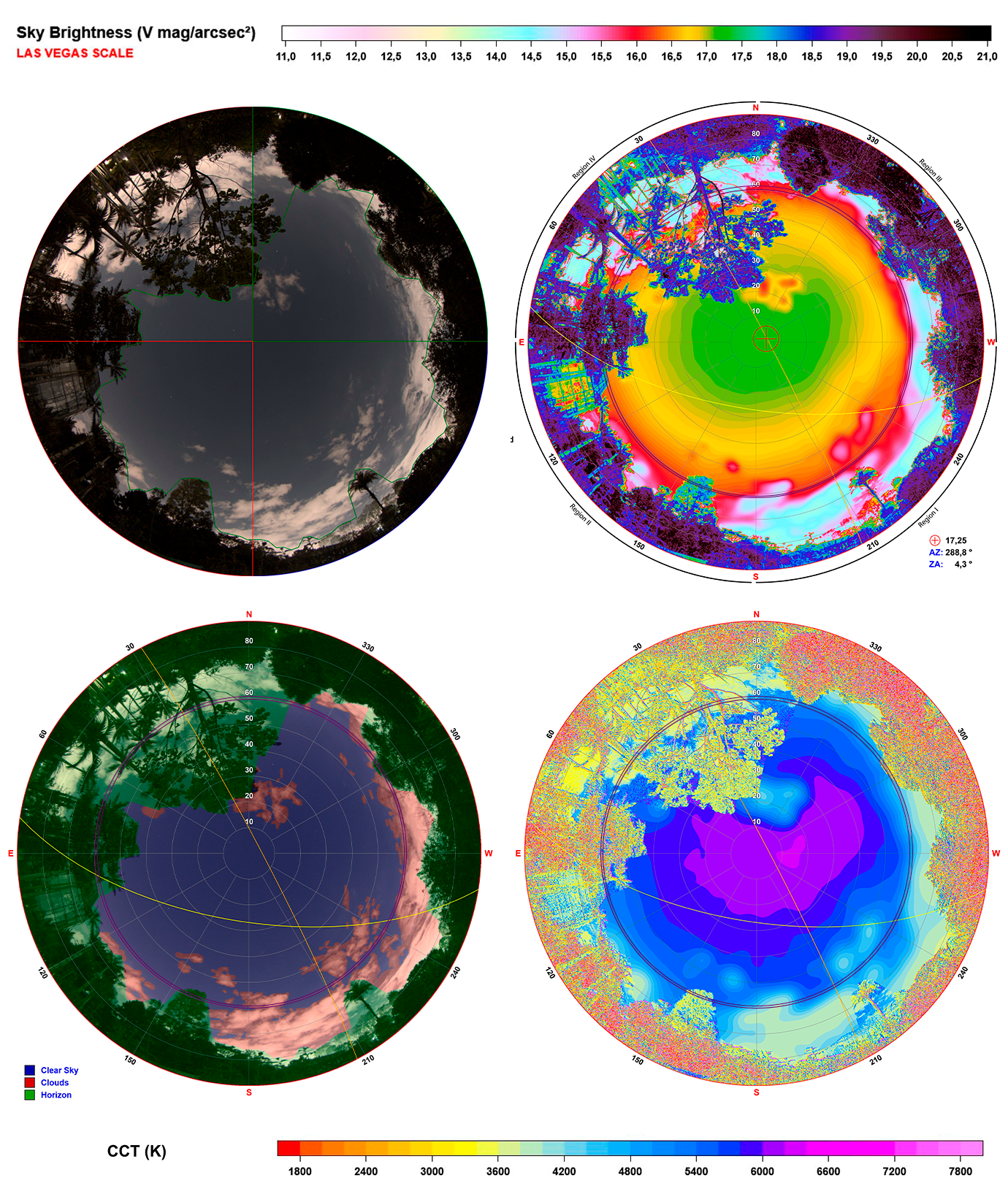}
    \caption{Same as Figure \ref{fig:sqc_OAUTP}, but for BBG on September 20, 2022. For more details, see Section \ref{Sec_JBB} and Table \ref{Tab:Areas}. It is important to note the distinct color scales used in the NSB and CCT maps, designed to accommodate the altered range of brightness and CCT levels characteristic of this heavily light-polluted urban sky.}
    \label{fig:sqc_JBB}
\end{figure*}

Figure \ref{fig:sqc_JBB} shows the all-sky view in RGB format from the BBG, highlighting the distribution and presence of clouds. The silhouettes of trees and part of the tropicarium structure are visible, with clouds distinctly appearing both amidst and above the treetops. Since the BBG is located in an urban area like Bogot\'a, where elevated levels of NSB are anticipated, the scales for the NSB and CCT maps were adjusted accordingly. The NSB map utilizes the Las Vegas scale, ranging from 11 to 21 V $mag/arcsec^2$, while the CCT map spans from 1600 to 7800 K. Both maps indicate an increased contribution from a $ZA=60^{\circ}$ towards the horizon.

\begin{figure*}[!t]
  \includegraphics[width=0.45\textwidth]{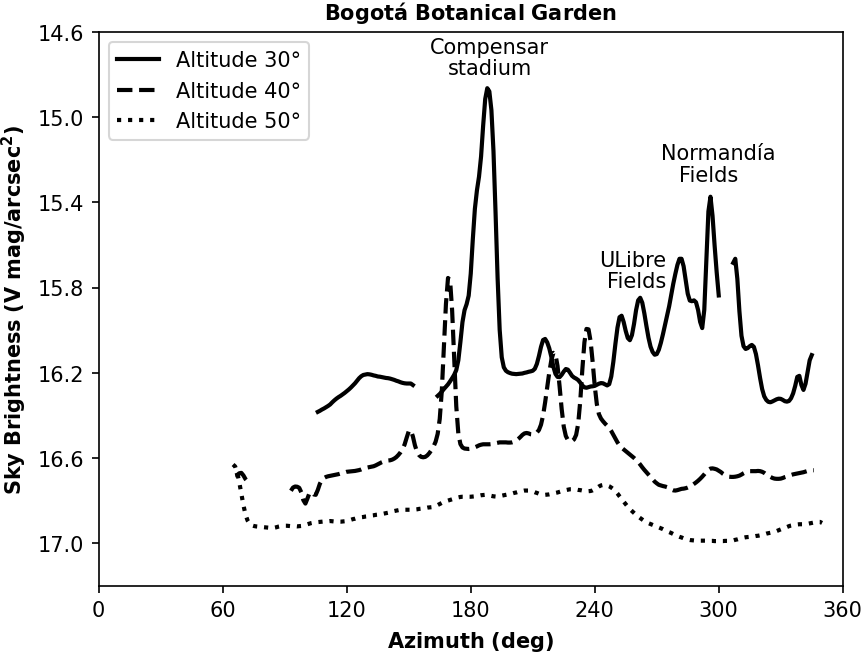}%
  \hspace*{\columnsep}%
  \includegraphics[width=0.45\textwidth]{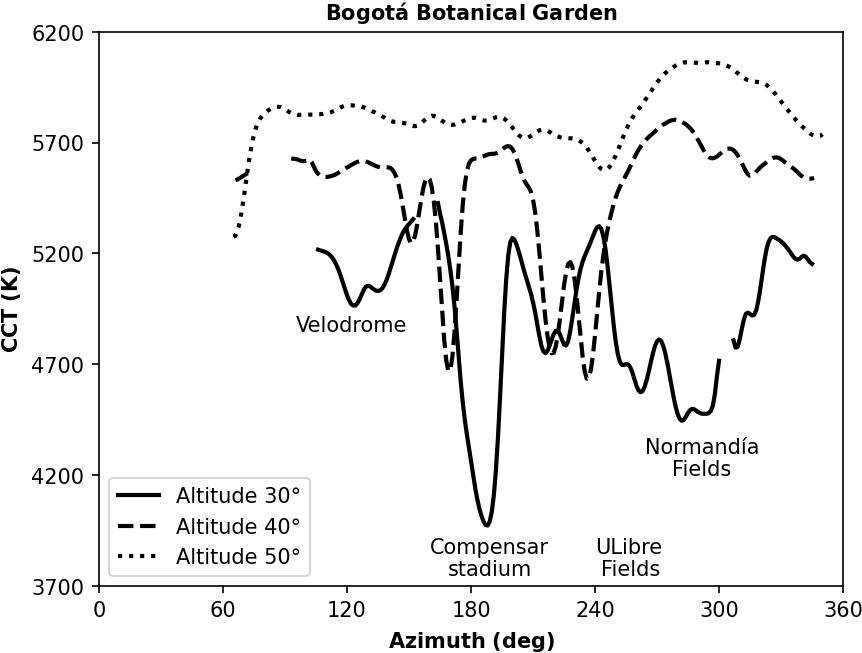}
  \caption{Same as Figure \ref{fig:UTP_Horizontal}, but for BBG. In this case, the rings are centered at $30^{\circ}$, $40^{\circ}$, and $50^{\circ}$ elevation. See Figure \ref{fig:sqc_JBB} and Table \ref{Tab:Areas}.}
  \label{fig:JBB_Horizontal}
\end{figure*}

To more accurately assess each region and identify the azimuths of significant contribution in NSB and CCT, Figure \ref{fig:JBB_Horizontal} illustrates the azimuthal profiles for the BBG. Selected altitudes of 30\degree, 40\degree, and 50\degree \ were chosen to mitigate interference from trees. However, Region IV is excluded from the profile due to the tropicarium and trees appearing prominently in that direction. Region I show a peak NSB and a minimum CCT at $AZ=188^{\circ}$, corresponding to the Compensar Stadium. Region II reveals its maximum NSB at $AZ=170^{\circ}$, associated with the bowling alley and auxiliary courts of Compensar, while the minimum CCT at $AZ=122^{\circ}$ aligns with the Velodrome. Region III displays the maximum NSB and minimum CCT at $AZ=296^{\circ}$, corresponding to the Normand\'ia fields.

\begin{figure*}[!t]
  \includegraphics[width=0.45\textwidth]{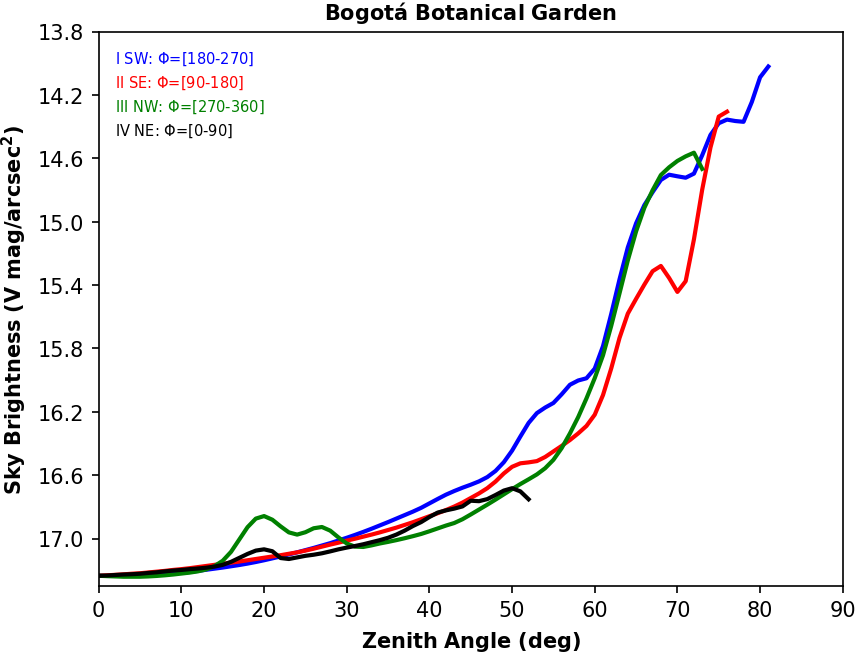}%
  \hspace*{\columnsep}%
  \includegraphics[width=0.45\textwidth]{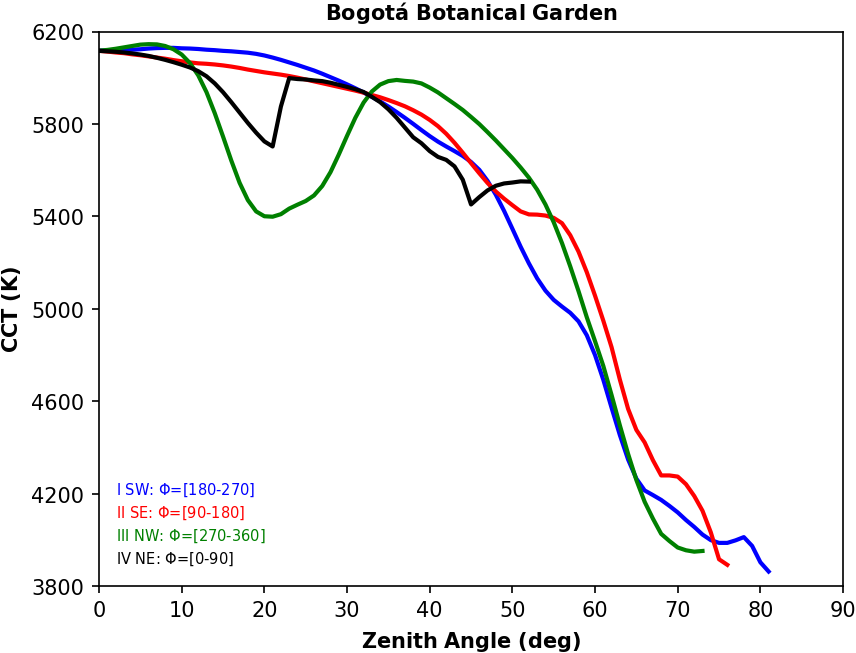}
  \caption{Same as Figure \ref{fig:UTP_Vertical}, but for BBG. For more details, see Section \ref{Sec_JBB}, Figure \ref{fig:sqc_JBB}, and Table \ref{Tab:Areas}.}
  \label{fig:JBB_Vertical}
\end{figure*}

The zenith profile is presented in Figure \ref{fig:JBB_Vertical}. Regions I and II exhibit an increase in NSB at $ZA=48^{\circ}$. Region III shows a rise in NSB from $ZA=14^{\circ}$ to $ZA=31^{\circ}$, attributable to the presence of a cloud, with a more pronounced increase from $ZA=55^{\circ}$. In Region IV, there is a gradual increase in NSB; however, data beyond $ZA=52^{\circ}$ is unavailable due to obstacles. Conversely, the zenith profile of CCT demonstrates a decrease as it approaches the horizon. In Region I, this decrease initiates at $ZA=39^{\circ}$. In Region II, the change occurs at $ZA=45^{\circ}$. Region III shows an initial decrease from $ZA=9^{\circ}$ to $ZA=34^{\circ}$, followed by a rapid decline beginning at $ZA=40^{\circ}$. Region IV exhibits a decrease from \textbf{$ZA=12^{\circ}$} to $ZA=23^{\circ}$, which is attributed to the reflection of light on the clouds. This decrease continues gradually until $ZA=52^{\circ}$.

\subsection{Cerro Guadalupe (CG)} \label{Sec_CG}

The Cerro de Guadalupe is part of the Cerros Orientales, a group of mountains located to the east of Bogot\'a and the Cruz Verde pramo in the eastern mountain range. These hills are designated as a national forest reserve and their strategic location makes them a crucial ecological connector for the city, facilitating the interaction of diverse mountain ecosystems \citep{garzon2014educacion}. The Cerros Orientales also play a vital role in regulating the climate and providing water for Bogot\'a \citep{bohorquez2008arriba}. Additionally, they boast rich and varied biodiversity, harboring numerous species of endemic flora and fauna that contribute to the conservation of the region's natural heritage \citep{jimenez2017ciudades}.

With an elevation of 3,273 meters above sea level, the Cerro de Guadalupe is the highest viewpoint accessible in Bogot\'a, offering a spectacular panoramic view of the Colombian capital. Since pre-Hispanic times, this site has held great significance, as rulers used it for surveillance and control over the city. Today, the hill is a subject of research in various fields such as history \citep{del2006monserrate}, environment \citep{rodriguez2015plan}, architecture \citep{meza2008urbanizacion} and archaeoastronomy (\citealt{romero2011aproximaciones}, \citealt{romeroarqueoastronomia}). It is also a prominent site of religious pilgrimage, housing the statue of the Virgin of Guadalupe, a place of worship for many faithful. The hill features hiking trails that attract tourists and mountaineering enthusiasts, who appreciate its natural beauty and unparalleled views \citep{esteban2017gestion}.

\begin{figure*}
    \centering
    \includegraphics[width=\textwidth]{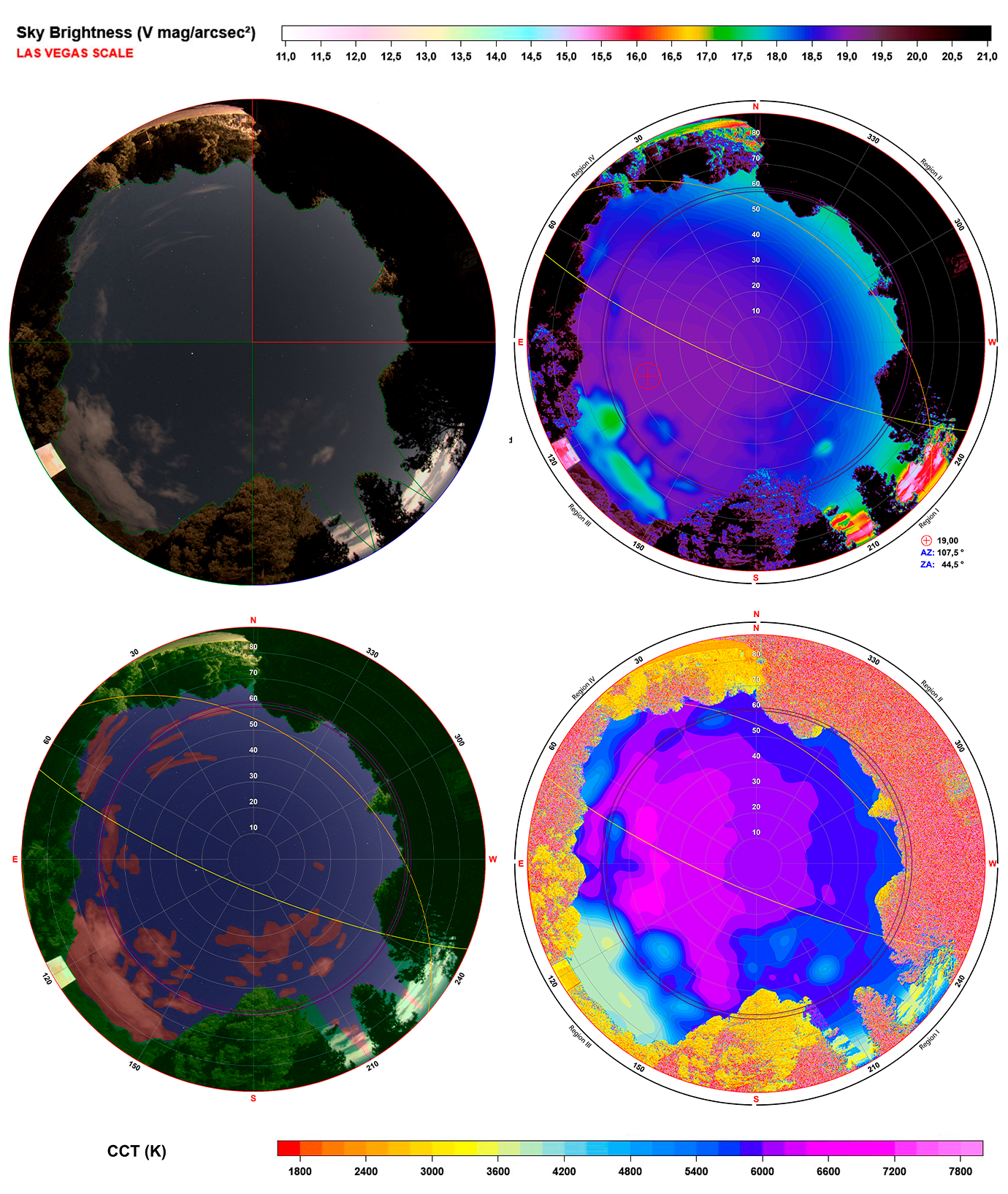}
    \caption{Same as Figure \ref{fig:sqc_JBB}, but for CG on September 20, 2022. For more details, see Section \ref{Sec_CG} and Table \ref{Tab:Areas}.}
    \label{fig:sqc_CG}
\end{figure*}

Figure \ref{fig:sqc_CG} presents all-sky view from the parking lot of the Sanctuary of the Virgin of Guadalupe on Cerro de Guadalupe. The RGB format in the top left panel showcases numerous trees and, to the southeast, the sanctuary's advertisement sign along with clouds in the same direction. Similar to the BBG, there is a noticeable scale change in NSB map and CCT map. Although this location is not within the city limits, its elevation above the city results in contributions from the urban environment. The NSB map (top right) indicates an increase in surface brightness between $AZ=240^{\circ}$ and $AZ=15^{\circ}$, attributed to the city of Bogot\'a. Additionally, there is a brightness increase from $AZ=90^{\circ}$ to $AZ=135^{\circ}$, corresponding to light scattered by clouds above the advertisement sign. This is corroborated by the cloud map and the CCT map (bottom right), which show a decrease in CCT in these regions.

\begin{figure*}[!t]
  \includegraphics[width=0.45\textwidth]{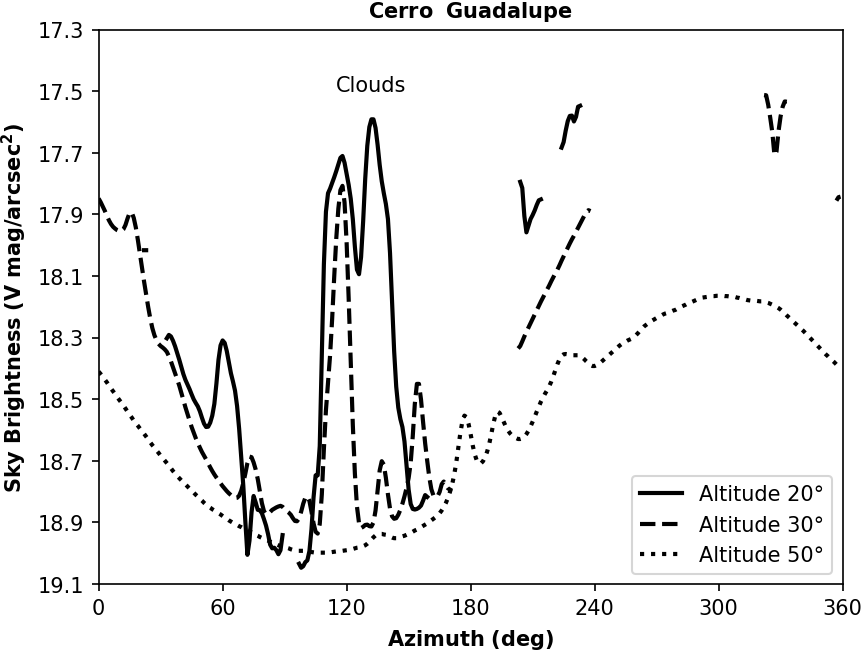}%
  \hspace*{\columnsep}%
  \includegraphics[width=0.45\textwidth]{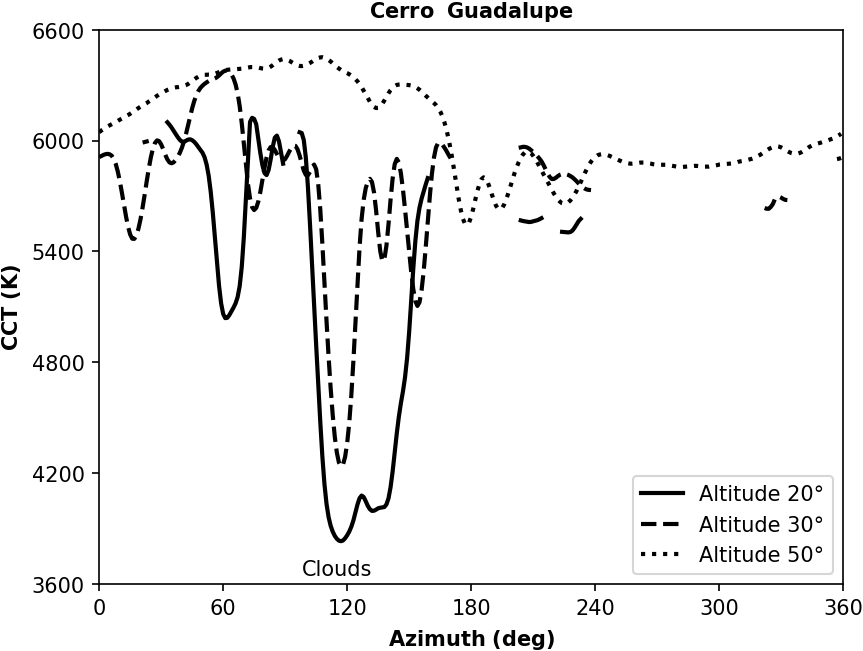}
  \caption{Same as Figure \ref{fig:UTP_Horizontal}, but for CG. In this case, the rings are centered at $20^{\circ}$, $30^{\circ}$, and $50^{\circ}$ elevation. See Figure \ref{fig:sqc_CG} and Table \ref{Tab:Areas}.}
  \label{fig:CG_Horizontal}
\end{figure*}

This decrease is more precisely detailed in the azimuthal profile. Figure \ref{fig:CG_Horizontal} displays rings centered at altitudes of $20^{\circ}$, $30^{\circ}$, and $50^{\circ}$, accounting for the presence of trees along the horizon.
The $50^{\circ}$ altitude ring is the only one that provides full azimuthal coverage, as vegetation is unevenly distributed. In contrast, the $20^{\circ}$ and $30^{\circ}$ altitude rings capture variations in NSB and CCT only in specific directions, limiting the ability to detect changes in these parameters across all azimuths. The most significant NSB peak is at $AZ=232^{\circ}$, corresponding to the city of Bogot\'a. Additionally, there are notable increases in surface brightness at $AZ=132^{\circ}$ and $AZ=117^{\circ}$. For CCT, two peaks indicate a decrease, corresponding to the city's glow on the cloud, with another decrease peak at $AZ=62^{\circ}$, corresponding to the direction of the church on the hill.

\begin{figure*}[!t]
  \includegraphics[width=0.45\textwidth]{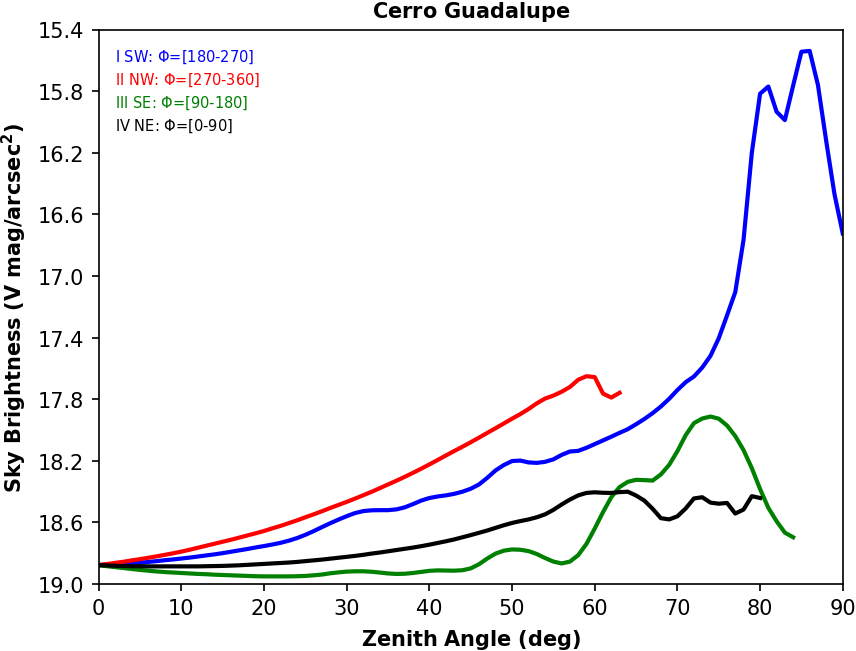}%
  \hspace*{\columnsep}%
  \includegraphics[width=0.45\textwidth]{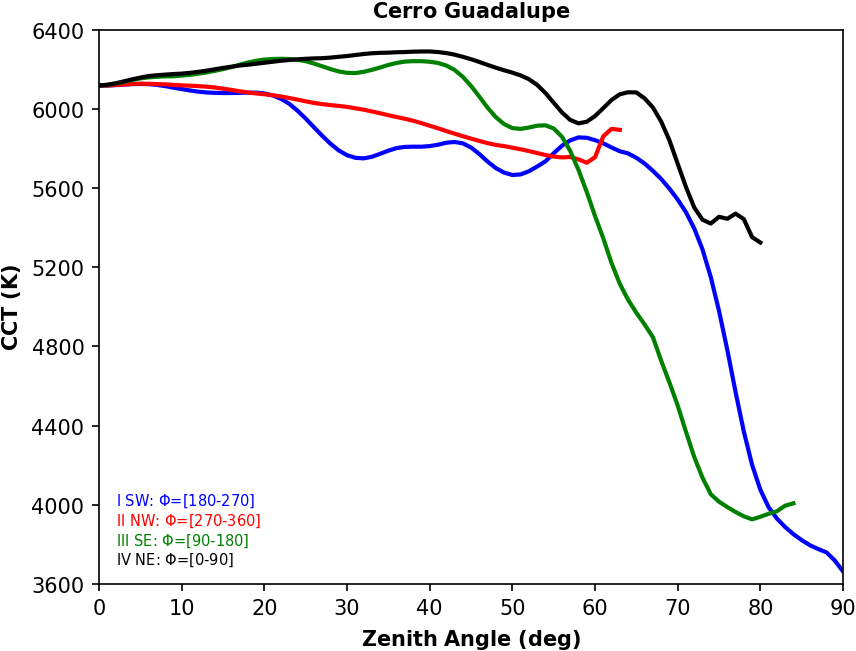}
  \caption{Same as Figure \ref{fig:UTP_Vertical}, but for CG. For more details, see Section \ref{Sec_CG}, Figure \ref{fig:sqc_CG}, and Table \ref{Tab:Areas}.}
  \label{fig:CG_Vertical}
\end{figure*}

The zenith profile is illustrated in Figure \ref{fig:CG_Vertical}. Region I shows a sharp increase in NSB at $ZA=72^{\circ}$, which decreases at $ZA=80^{\circ}$ due to light dispersion by the cloud and further decreases from $ZA=87^{\circ}$ due to the city's glow. The CCT decreases from $ZA=58^{\circ}$. Region II exhibits an increase in NSB and a decrease in CCT from the zenith to $ZA=60^{\circ}$, limited by tree coverage. Region III presents the most significant increase in NSB and decrease in CCT starting at $ZA=57^{\circ}$, corresponding to a cloud in the sector. Region IV shows a gradual increase from the zenith to $ZA=64^{\circ}$, reaching maximum brightness, then decreasing and increasing again due to clouds among the trees, with the strongest CCT decrease at $ZA=65^{\circ}$.

\section{Discussion}\label{Sec_4}

In this section, we compare our findings with those from similar sites worldwide and with available numerical models. We also explore the implications of these results for conserving and protecting the country's dark sky. The discussion includes possible measures to mitigate light pollution and preserve this unique natural resource. 
\subsection{Photometric indicators for cross-site comparisons}\label{Sec_4.1}

For comparative studies between different locations, a large number of photometric indicators of the visual quality of the night sky have been suggested in the literature (\citealt{deverchere2022towards}; \citealt{duriscoe2016photometric}). These indicators enable us to evaluate the impact of light pollution from various perspectives, such as astronomical \citep{falchi2023light} and biological \citep{falchi2021computing}. In Table \ref{Tab:Min-Max}, we present the alt-az coordinates and the corresponding NSB/CCT values for the darkest and brightest points of the sky, as well as for the zenith (averaged over a 5\degree \ radius circle), within a 1\degree \ wide ring centered at 30\degree \ elevation, and over the entire celestial hemisphere (i.e., all-sky) for each site.

\begin{table*}
\centering
  \newcommand{\DS}{\hspace{5\tabcolsep}} 
  \setlength{\tabnotewidth}{\textwidth}
  \setlength{\tabcolsep}{0.7\tabcolsep}
  \tablecols{15}
  \caption{Photometric indicators.\tabnotemark{a}} 
  \label{Tab:Min-Max}
  \resizebox{\textwidth}{!}{%
  \begin{tabular}{ccccccccccccccc}
    \toprule
    \textbf{Site} & \multicolumn{4}{c}{\textbf{Brightest Point\tabnotemark{$^{\alpha}$}}} & \multicolumn{4}{c}{\textbf{Darkest Point\tabnotemark{$^{\alpha}$}}} & \multicolumn{2}{c}{\textbf{Zenith\tabnotemark{$^{\beta}$}}} & \multicolumn{2}{c}{\textbf{30$^{\circ}\tabnotemark{$\gamma$}$}} & \multicolumn{2}{c}{\textbf{All-Sky}} \\ 
    \cmidrule{2-15} 
    \textbf{Acronym} & \textbf{AZ} & \textbf{ZA} & \textbf{NSB} & \textbf{CCT} & \textbf{AZ} & \textbf{ZA} & \textbf{NSB} & \textbf{CCT} & \textbf{NSB} & \textbf{CCT} & \textbf{NSB} & \textbf{CCT} & \textbf{NSB} & \textbf{CCT}\\
    \midrule
    AOUTP & 290 & 78 & 15.45 & 3434 & 160 & 8 & 18.12 & 4884 & 18.09 & 4860 & 17.11 & 4481 & 16.90 & 4407 \\
    \cmidrule{1-15} 
    TD & 206 & 90 & 17.41 & 4321 & 66 & 23 & 21.36 & 4041 & 21.29 & 4061 & 20.77 & 3995 & 20.59 & 4012 \\
    \cmidrule{1-15} 
    BBG & 194 & 71 & 14.21 & 3730 & 289 & 4 & 17.25 & 6181 & 17.23 & 6100 & 16.01 & 4908 & 16.09 & 4561 \\
    \cmidrule{1-15} 
    CG & 210 & 86 & 15.53 & 3503 & 107 & 44 & 19.00 & 6584 & 18.87 & 6131 & 18.39 & 5591 & 18.43 & 5502 \\
    \bottomrule
    \tabnotetext{a}{Altazimuth coordinates for the NSB (V mag/arcsec$^2$) and CCT (K) values of the darkest and brightest points in the sky at the time of our observations. Additionally, values for the zenith, a ring of 30$^{\circ}$ of elevation, and the entire celestial vault are included.}
    \tabnotetext{$^{\alpha}$}{Average value within 1 $deg^2$.}
    \tabnotetext{$^{\beta}$}{Average value for zenith angle $ZA=5^{\circ}$.}
    \tabnotetext{$^{\gamma}$}{ Average value at 30$^{\circ}$ elevation with a width of 1$^{\circ}$ (i.e., for $59.5^{\circ} < ZA < 60.5^{\circ}$).}
  \end{tabular}
}
\end{table*}

\begin{figure*}
    \centering
    \includegraphics[width=\textwidth]{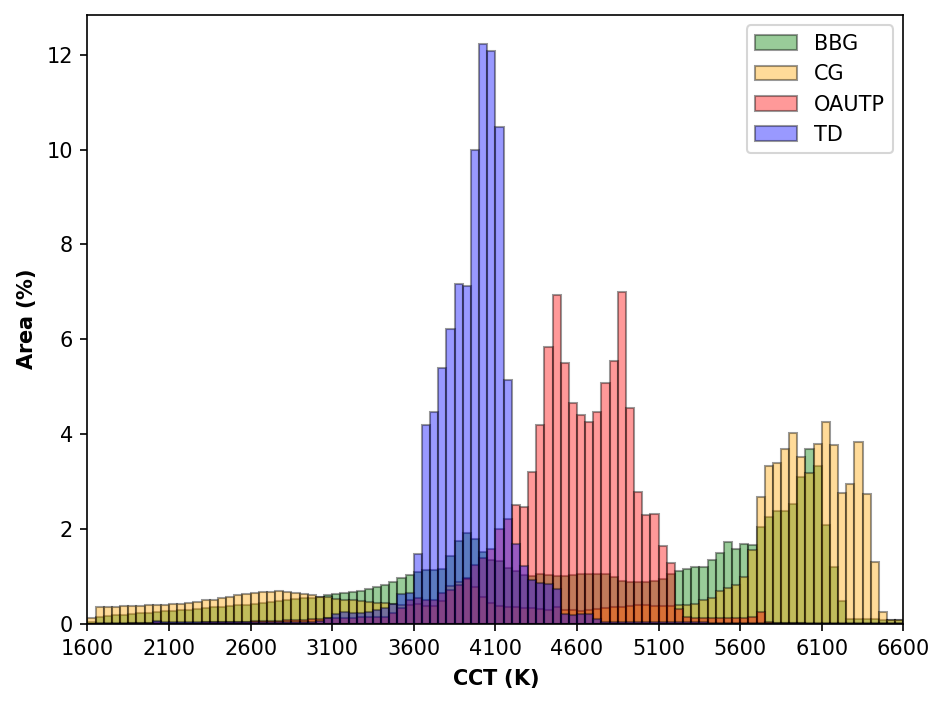}
    \caption{The histogram represents the distribution of CCT values for four measurement sites: BBG (green), CG (yellow), AOUTP (red), and TD (blue). Each color corresponds to a different site, showing the proportion of area (\%) covered by different CCT values across the range of 1600 to 6600 K. The bin width is set to 100 K}
    \label{fig:Hist}
\end{figure*}

To better understand the types of light affecting the sky at each site, Figure \ref{fig:Hist} presents a comparative histogram illustrating the percentage distribution of the sky based on CCT. The histogram for TD, shown in blue, reflects a distribution with temperatures typical of a natural sky, but it also reveals a significant percentage of LED light, with a peak around 4100 K, indicating the presence of neutral-tone LEDs.

In contrast, the histogram for AOUTP, represented in red, shows a bimodal distribution with peaks at 4500 K and 5000 K, suggesting a sky influenced by urban light sources. The 4500 K peak is associated with neutral light LEDs, commonly used in commercial areas, while the 5000 K peak corresponds to cool light LEDs, typical of large installations such as sports arenas, parking lots, or industrial facilities. Lastly, the histograms for BBG and CG reveal significant concentrations in the higher CCT ranges, between 5000 K and 6500 K. The peak near 6100 K suggests the presence of cool or high-intensity LEDs, characteristic of densely illuminated urban areas like Bogot\'a.

\subsection{Comparison with Other Sites}

The NSB and CCT maps of the Colombian skies presented in this study, derived from data collected by the SQC, constitute the first recorded measurements of light pollution near the equator as documented in the Light Pollution Map database. This platform compiles all files generated by the SQC instrument, allowing users to upload and share their observations, thereby ensuring public access to the data. Despite significant efforts to facilitate access to SQC data, measurements are still lacking for many cities.

For this reason, when comparing our data with those from the database, specific selection criteria were applied for each location. The AOUTP was compared with the Astronomsko-geofizikalni Observatorij Golovec (AOG) in Slovenia, as both are observatories located less than 10 km from the center of their respective cities. For TD, the Eifel National Park\footnote{https://www.eifel.info/} (ENP) was chosen, as both sites are designated for dark sky preservation. In the case of BBG and CG, a comparison was made with Mont Royal Park\footnote{https://montreal.ca/en/places/parc-du-mont-royal} (MRP) and the Montreal Olympic Stadium (MOS), both located in Montreal, Canada. This comparison was based on the search for a city that had two measurement points no more than 12 km apart, where one of the sites was a park or an environmentally protected area.

First, the AOUTP was compared with the AOG in Slovenia. The AOG is located in Ljubljana, a city of approximately 285,604 inhabitants (2021 data\footnote{https://www.stat.si/}), while the AOUTP is situated in Pereira, which has around 477,068 inhabitants (2018 data\footnote{https://www.dane.gov.co/}). They exhibit significant differences in terrain and climate, which may influence light pollution measurements. Ljubljana lies in a basin surrounded by mountains, potentially trapping light pollution, whereas Pereira is located in the Andean region, with distinct atmospheric conditions. These differences highlight the importance of contextual factors when interpreting light pollution data.

\begin{figure*}
    \centering
    \includegraphics[width=\textwidth]{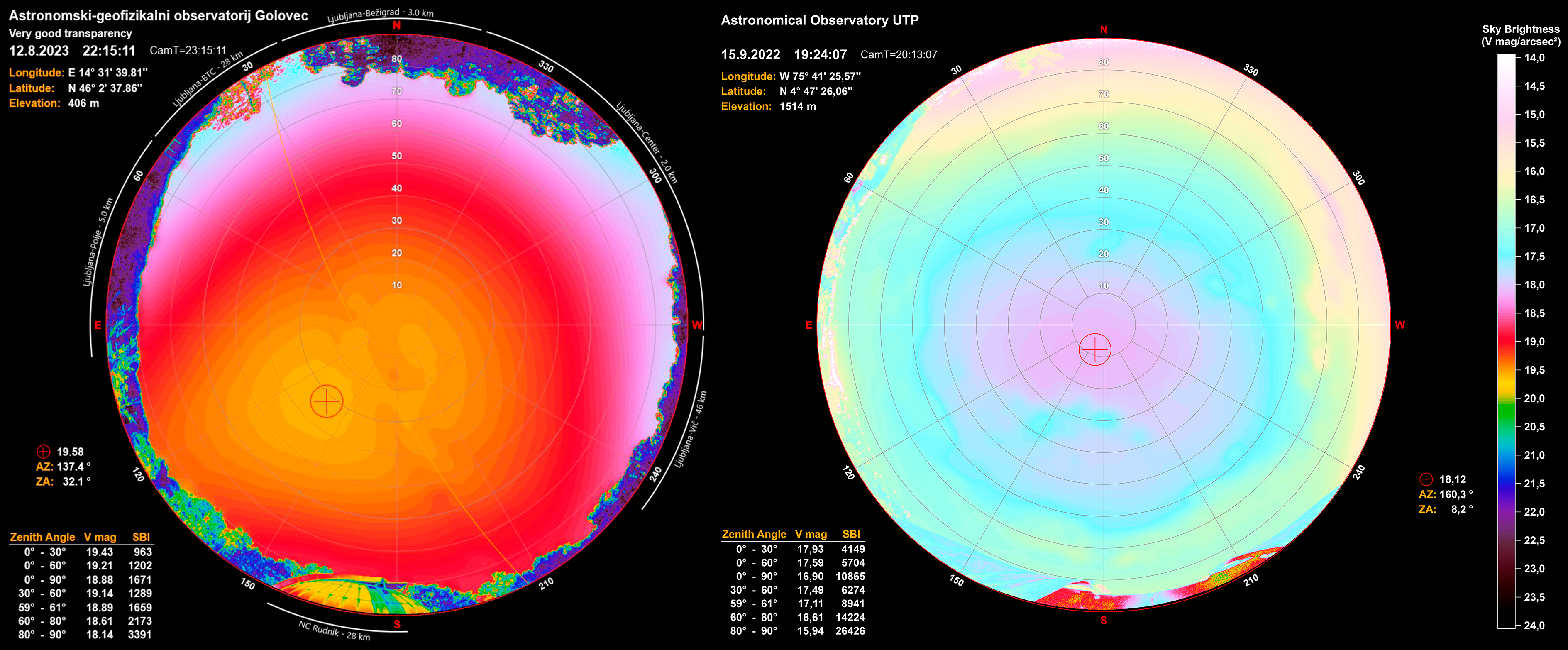}
    \caption{Comparison of all-sky NSB images from the AOG (left) obtained from the Light Pollution database and from the AOUTP (right). The date, time, and geographic coordinates are displayed at the top left of each map, while NSB values in $[V mag/arcsec^2]$ are represented in the color bar on the right. The dark point, along with its coordinates, is marked above the table containing the NSB data at the bottom left.}
    \label{fig:UTP_Comp}
\end{figure*}

Figure \ref{fig:UTP_Comp} compares the NSB maps of AOUTP and AOG. From the maps, the darkest point values are highlighted, which are 18.12 $mag/arcsec^2$ and 19.58 $mag/arcsec^2$, respectively. The photometric indicator for the 30\degree \ elevation ring shows values of 17.11 $mag/arcsec^2$ and 18.89 $mag/arcsec^2$, respectively. This indicates that the night sky over AOG is 1.78 $mag/arcsec^2$ darker compared to Pereira. This difference can be attributed to several factors, including the higher population density of Pereira and the presence of an airport within its urban area, both of which contribute to increased NSB. Moreover, the frequent partial cloud cover over Pereira further elevates the recorded NSB by scattering and amplifying artificial light sources.

In the case of TD, we compared our data with ENP, located in western Germany within the state of North Rhine-Westphalia, near the Belgian border in the Eifel mountain region. The park is situated close to the towns of Schleiden and Monschau, as well as the cities of Aachen and Cologne. Covering approximately $110 \ km^2$, ENP is home to thousands of endangered animal and plant species. In 2014, it was designated as a protected area by the International Dark Sky Association\footnote{https://www.darksky.org/}, becoming one of the few regions in Germany where the Milky Way is visible to the naked eye \citep{gabriel2017resources}.

\begin{figure*}
    \centering
    \includegraphics[width=\textwidth]{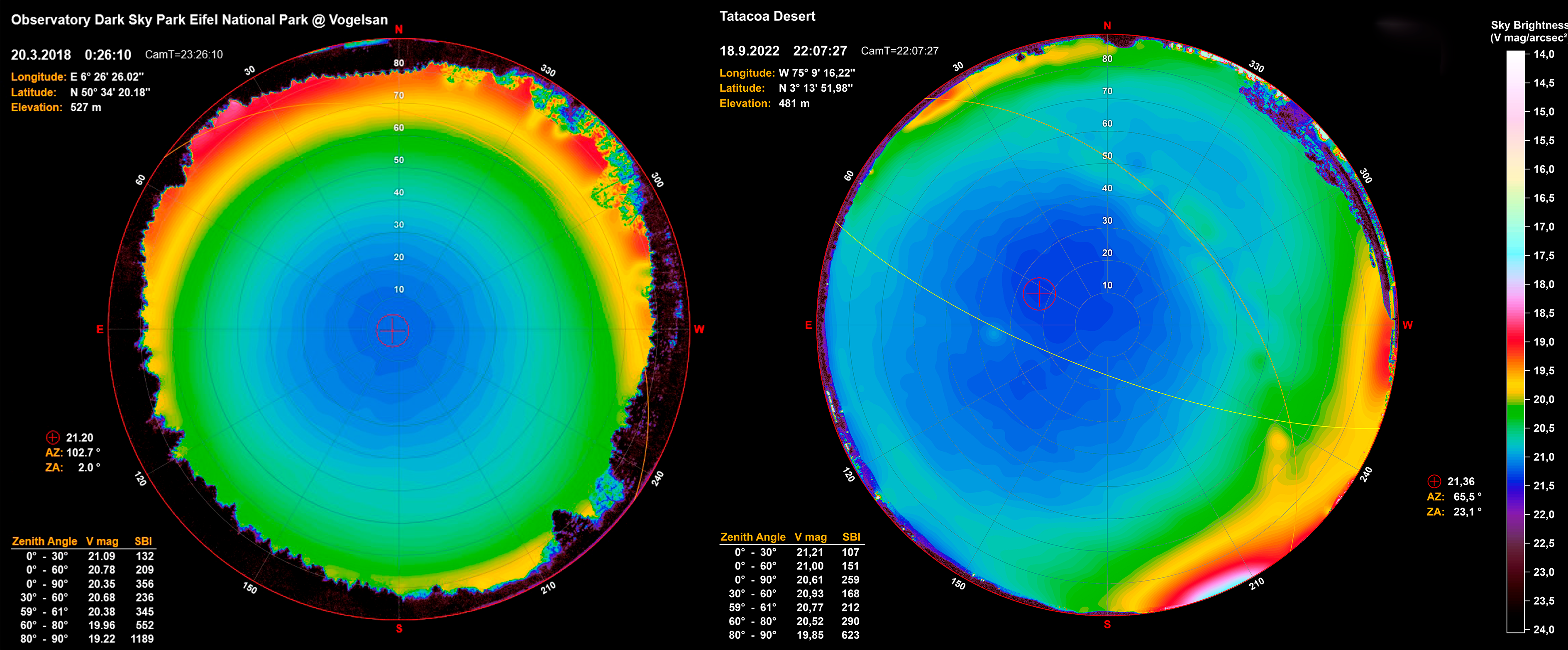}
    \caption{Comparison of ENP and TD data, with analogous information to Figure \ref{fig:UTP_Comp}.}
    \label{fig:DT_Comp}
\end{figure*}

Figure \ref{fig:DT_Comp} the NSB maps of TD and ENP. From the maps, the values of the darkest points stand out, which are 21.36 $mag/arcsec^2$ and 21.22 $mag/arcsec^2$, respectively. The photometric indicator for the 30\degree \ elevation ring shows values of 20.77 $mag/arcsec^2$ for TD and 20.38 $mag/arcsec^2$ for ENP, indicating that the night sky over TD is 0.39 $mag/arcsec^2$ darker than that over ENP. This difference is primarily due to the proximity of large cities near ENP, such as Cologne (Germany, 50 km away) and Li\'ege (Belgium, 60 km away), compared to Neiva, which is 37 km away from TD. Furthermore, the populations of Cologne and Li\'ege are 2.7 and 1.7 times larger than that of Neiva, respectively, contributing to the higher NSB observed in ENP. Despite these differences, it is important to note that both sites have implemented measures to protect their night skies, supported by their respective certifications. However, special attention should be given to the NSB levels at TD, as ENP is located in one of the most densely populated regions in Europe, making it particularly vulnerable to the effects of surrounding light pollution.

As outlined earlier, we selected two locations in each city, situated no more than 12 km apart, for analysis. The first point of interest focuses on conservation parks that blend natural and urban environments: BBG in Bogotá and MRP in Montreal. These spaces are distinguished by their commitment to environmental preservation within an urban setting.

The second comparison point includes CG in Bogotá and MOS in Montreal. Both locations are within 12 km of their respective parks and serve as prominent city viewpoints. In Bogotá, CG is a religious sanctuary that, beyond its spiritual significance, provides a breathtaking panoramic view of the city. Meanwhile, MOS features the iconic inclined tower of the Montreal Olympic Stadium, the tallest of its kind in the world. Known as the Montreal Tower, it offers a stunning 360-degree vantage point, allowing visitors to admire the city and its surroundings from an unparalleled perspective.

These cities exhibit notable differences in their geography and demographics. Bogotá and Montreal exhibit notable geographic and demographic differences. Bogotá, situated high in the Andes Mountains at 2,565 meters above sea level, is defined by its rugged, mountainous terrain. In contrast, Montreal is located on an island in the St. Lawrence River, featuring a relatively flat and gently rolling landscape. Demographically, Bogotá is significantly larger, with approximately 8.034.649 inhabitants as of 2018 according to DANE, compared to Montreal's 1,762,949\footnote{https://www.statcan.gc.ca/} residents in 2021. However, Montreal is notable for having multiple SQC measurements recorded in the Light Pollution Database.

\begin{figure*}
    \centering
    \includegraphics[width=\textwidth]{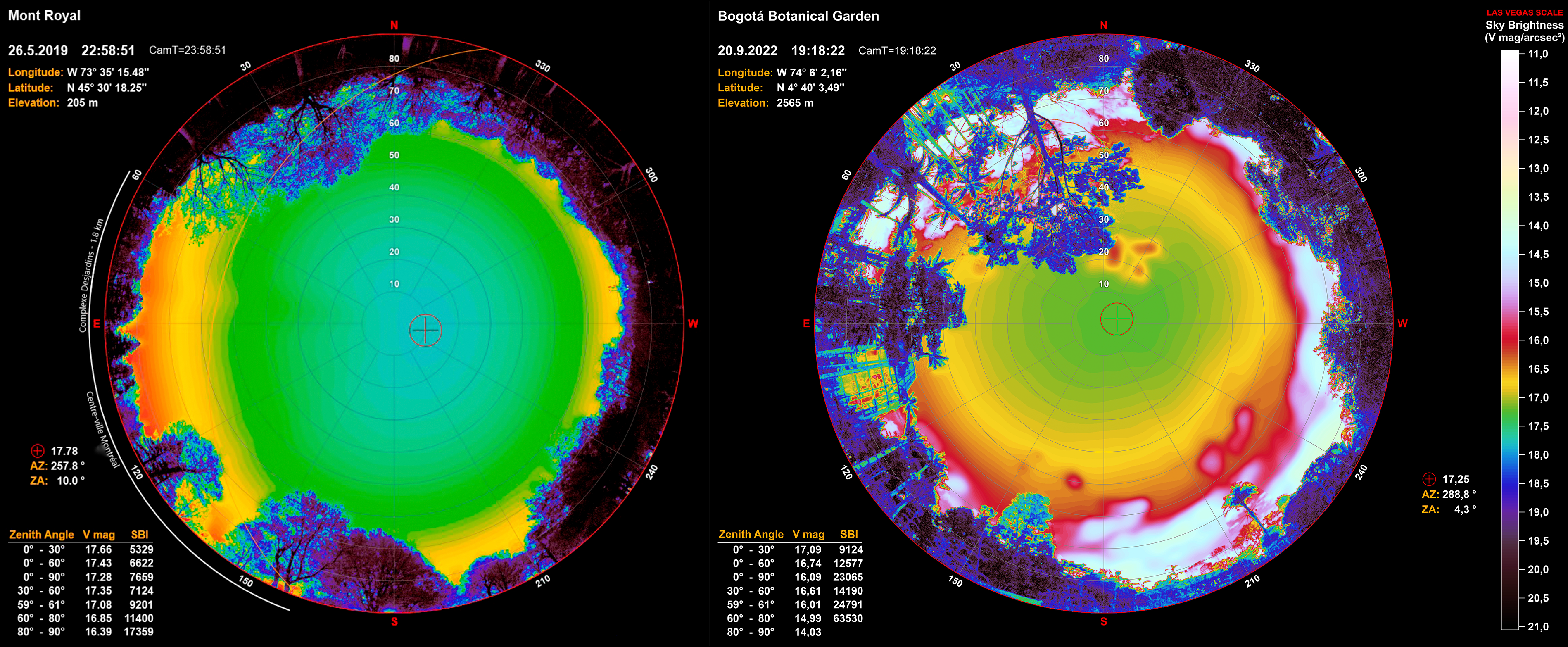}
    \caption{Comparison of MRP and BBG data, with analogous information to Figure \ref{fig:UTP_Comp}.}
    \label{fig:JBB_Comp}
\end{figure*}

Figure \ref{fig:JBB_Comp} compares the NSB maps of MRP and BBG, highlighting the darkest points, which are 17.78 $mag/arcsec^2$ and 17.25 $mag/arcsec^2$, respectively. The photometric indicator (30\degree \ elevation ring) shows values of 17.08 $mag/arcsec^2$ and 16.01 $mag/arcsec^2$, respectively. According to the photometric indicator, MRP has a sky approximately 1.07 $mag/arcsec^2$ darker than BBG. This disparity is likely influenced by several factors, including Bogot\'a's size, which is 4.4 times that of Montreal.

\begin{figure*}
    \centering
    \includegraphics[width=\textwidth]{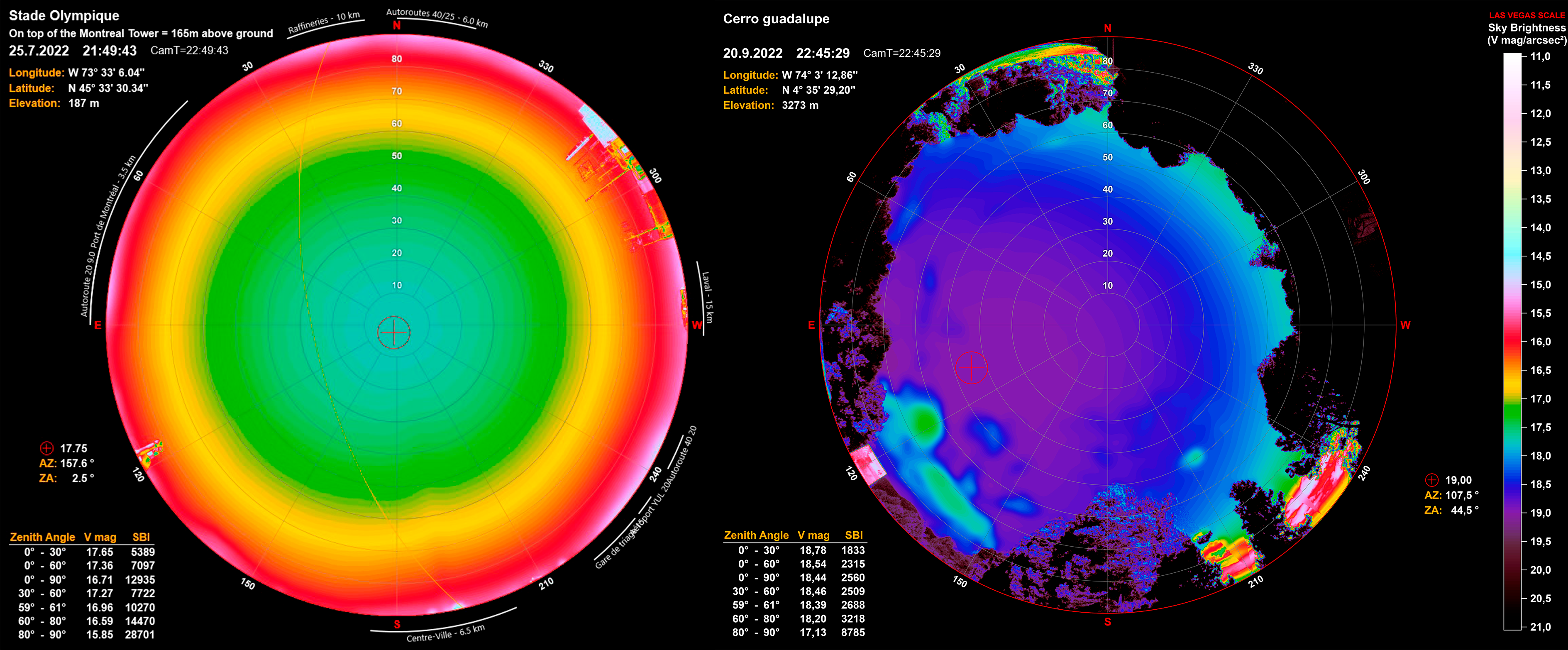}
    \caption{Comparison of MOS and CG data, with analogous information to Figure \ref{fig:UTP_Comp}.}
    \label{fig:CG_Comp}
\end{figure*}

Figure \ref{fig:CG_Comp} compares the NSB maps of MOS and CG, highlighting the values of the darkest points, which are 17.75 $mag/arcsec^2$ and 19.00 $mag/arcsec^2$, respectively. The photometric indicator (30\degree \ elevation ring) shows values of 17.27 $mag/arcsec^2$ and 18.39 $mag/arcsec^2$, respectively. According to this indicator, CG presents a sky approximately 1.12 $mag/arcsec^2$ darker than MOS. This difference could be attributed to the location of CG on the outskirts of Bogot\'a, in contrast to MOS, which is situated within the city's interior. Both locations provide panoramic views of the city: CG, situated at an elevation of approximately 3,273 meters above sea level, and MOS, at 167 meters. From MOS, one can observe the city's most iconic buildings and urban landscape, while CG serves as a natural lookout that also offers opportunities to observe local wildlife.

\subsection{GAMBONS Model}\label{Sec_GAMBONS}

The night sky is not completely dark due to contributions from several natural sources, including integrated starlight, diffuse galactic light, extragalactic light, and zodiacal light. Additionally, phenomena such as airglow result from the interaction of particles in the upper atmosphere. These natural effects create a background glow that is always present in the night sky.

The GAMBONS model, as detailed in \citet{masana2021multiband, masana2022enhanced}, accounts for various factors to simulate natural NSB under conditions devoid of clouds, the Moon, and planets. A key innovation of GAMBONS is its seamless integration of the Gaia-DR3 and Hipparcos catalogs, combined with detailed modeling of diffuse galactic and extragalactic light, zodiacal light, airglow, and atmospheric attenuation and scattering effects. The model operates with default airglow and aerosol parameters, providing a reliable initial approximation of the typical conditions at our monitoring sites.

Using these parameters, GAMBONS generates an all-sky NSB map for a specified time and location. The presence or absence of the Milky Way and the elevation of the ecliptic above the horizon are the primary factors influencing this model. Additionally, the model can estimate synthetic NSB values in different photometric passbands, such as the SQM, TESS, or specific photometric bands. This feature is precious for theoretically assessing natural light contributions and comparing them with observed data, thereby facilitating the estimation of the percentage of ALAN at each site.

Table \ref{Tab:Gambons} compares values measured with the SQC at the zenith and all-sky, alongside the values obtained from the GAMBONS model (GMB). The final column indicates the estimated percentage of ALAN at each location. This estimation is derived from the model's values, with conversions following the methodology outlined in \citep{kollath2020introducing}. It is important to note that the GMB model tends to overestimate natural NSB near the horizon and underestimate it at the zenith by up to 0.1 $mag/arcsec^2$ \citep{masana2022enhanced}.

\begin{table}
\centering
  \newcommand{\DS}{\hspace{7\tabcolsep}} 
  \setlength{\tabnotewidth}{\columnwidth}
  \setlength{\tabcolsep}{0.35\tabcolsep}
  \tablecols{7}
  \caption{Comparison of NSB SQC vs GMB.\tabnotemark{a}} 
  \label{Tab:Gambons}
  \begin{tabular}{c c c c c c c }
    \toprule
    \textbf{Site}  & \multicolumn{3}{c}{\begin{tabular}[c]{@{}c@{}}\textbf{Zenith}\end{tabular}} & \multicolumn{3}{c}{\begin{tabular}[c]{@{}c@{}}\textbf{All-sky}\end{tabular}} \\
    \cmidrule{2-4}
    \cmidrule{5-7}
    \textbf{acronym} & \textbf{GMB} & \textbf{SQC} & \textbf{$\Delta m$} & \textbf{GMB} &\textbf{SQC} & \textbf{ALAN$\%$} \\ 
    \midrule
    AOUTP & 21.46 & 18.09 &  3.37 & 21.45 & 16.90 & 98    \\
    \cmidrule{1-7}
    TD  & 21.57 & 21.29  & 0.28 & 21.56 & 20.59 & 59   \\  
    \cmidrule{1-7}
    BBG &  21.26 & 17.23  & 4.03  & 21.43 & 16.08 & 99  \\  
    \cmidrule{1-7}
    CG  & 21.60 & 18.87  &2.73 & 21.53 & 18.43 & 94  \\ 
    \bottomrule
    \tabnotetext{a}{Comparison between the GAMBONS model (GMB) and the observed NSB (SQC) at  zenith (V $mag/arcsec^2$) and all-sky. For more details, see Section \ref{Sec_GAMBONS}.}
  \end{tabular}
\end{table}

TD and BBG are the only sites for which prior ALAN data were available in the literature, all recorded using the SQM photometer (21.26 $mag/arcsec^2$ for TD and 16.31 $mag/arcsec^2$ for BBG). These historical data are utilized here for a direct comparison with our SQC measurements, providing a potential means to estimate the evolution of ALAN over the past years (\citealt{urrego2016analisis} (hereafter UG, 2016); \citealt{goez2021comparative} (hereafter GT\&VD, 2021)). 

Table \ref{tab:GambonsRef} compares the values reported in the literature (SQM LIT) with those obtained using the GMB models (SQM GMB) for the same location, date, and time. While direct measurements taken with a SQM are not available in this study, the SQC methodology enables the calculation of a synthetic SQM value based on SQC data. This synthetic value is included in the SQM LIT column. Conversely, the SQM GMB column reports the results of the GMB model simulations for the dates and times analyzed in this study.

Although a direct comparison is challenging due to variations in dates, times, measurement locations, instruments used, and meteorological conditions, the obtained values provide valuable insights into the evolution of NSB at both locations. To account for intrinsic and previously unknown variations in natural conditions, the original SQM measurements reported in the literature and our synthetic values derived from SQC were compared with the theoretical predictions generated by the GMB SQM model.

For this purpose, a column labeled $\Delta m$ is included, representing the difference in NSB between the SQM GMB and SQM LIT values. Similarly, the corresponding row includes $\Delta M$, which indicates the time difference between the SQM GMB and SQM LIT values. These metrics not only facilitate the evaluation of discrepancies and similarities between the simulated data and the literature-reported data but also enable an analysis of the temporal evolution of NSB.

\begin{figure*}[!t]
    \centering
    \includegraphics[width=\textwidth]{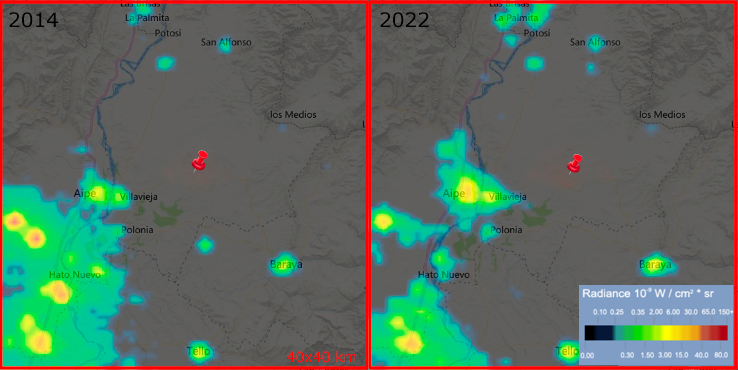}
    \caption{Comparison of NSB using VIIRS/DNB Radiance data in a 40x40 km area centered on the 2022 measurement point at TD (red pin) for the years 2014 (left) and 2022 (right).}
    \label{TDVIIRS}
\end{figure*}

\begin{figure*}[!t]
    \centering
    \includegraphics[width=\textwidth]{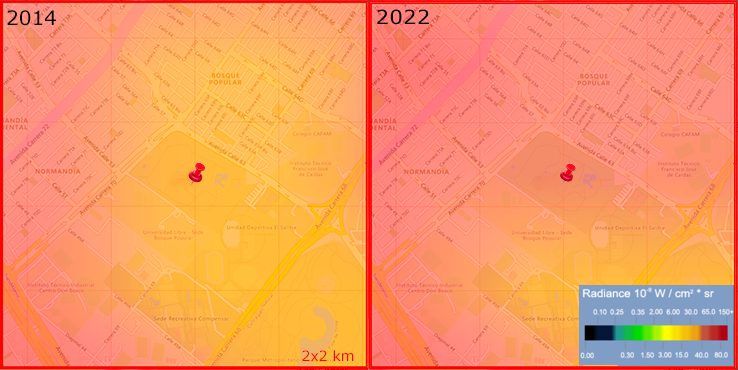}
    \caption{Same as Figure \ref{TDVIIRS}, \textbf{but} focused on BBG and centered within a 2x2 km area.}
    \label{BGVIIRS}
\end{figure*}

Figures \ref{TDVIIRS} and \ref{BGVIIRS}, generated using VIIRS Day/Night Band  (DNB) data sourced from the Light Pollution Map platform, illustrate the temporal evolution of NSB for the years 2014 and 2022, respectively. Figure \ref{TDVIIRS} covers an area of \textbf{1600 $km^2$} and shows a radiance of $2840.6 \cdot 10^{-9}$ $ W/cm^{2} \cdot sr$ in 2014, decreasing to $1595.3 \cdot 10^{-9}$ $ W/cm^{2} \cdot sr$ in 2022. This suggests a reduction in the way the desert is being illuminated, which can be visually observed as a decrease in radiance in the lower-left part of the image. However, this area corresponds to the Dina Field, one of the oil and gas production fields operated by Ecopetrol, the main oil and gas company in Colombia. 
This observed reduction may be attributed to the adoption of LED technology, the predominant lighting source during the measurement period. However, Table \ref{tab:GambonsRef} reveals an increase in NSB, whereas VIIRS data suggest an apparent decrease. This discrepancy is primarily due to the limited sensitivity of the VIIRS DNB channel, which excludes wavelengths predominantly emitted by LEDs. Consequently, NSB tends to be underestimated in regions dominated by LED lighting. For a comprehensive discussion on the characteristics and limitations of VIIRS, refer to \citealt{cao2014quantitative}.

On the other hand, Figure \ref{BGVIIRS} covers an area of 4 $km^2$ and shows an increase in radiance, from $796.0 \cdot 10^{-9}$ $ W/cm^{2} \cdot sr$ in 2014 to $905.3 \cdot 10^{-9}$ $ W/cm^{2} \cdot sr$ in 2022. The increase in NSB is attributed not only to the 558 lamps\footnote{\url{https://datosabiertos.bogota.gov.co/dataset/luminarias_upz-bogota-d-c}} within the BBG, but also to the contributions from all the surrounding neighborhoods.

\begin{table}
\centering
  \setlength{\tabnotewidth}{\columnwidth}
  \setlength{\tabcolsep}{0.5\tabcolsep}
  \tablecols{4}
  \caption{Comparison of SQM Values } \label{tab:GambonsRef}
 \begin{tabular}{c c c c}
    \toprule
     \textbf{Date} & \textbf{SQM} & \textbf{SQM} & \multirow{2}{*}{\textbf{$\Delta m$}}\\
     \textit{aaaa-mm-dd}& \textbf{GMB} & \textbf{LIT} &  \\ 
    \midrule
    \multicolumn{4}{c}{TD} \\
    \cmidrule{1-4}
    2014-04-15 & 21.87 & 21.26 (GT\&VD, 2021) & 0.61 \\
    2022-09-18 & 21.81 & 21.09 (This work) & 0.72 \\ 
    \cdashline{1-4}
    \textbf{$\Delta M$} & 0.06 & 0.17 & - \\ 
    \cmidrule{1-4}
    \multicolumn{4}{c}{BBG} \\
    \cmidrule{1-4}
    2014-09-20 & 21.60 & 16.31 (UG, 2016)  & 5.29   \\
    2022-09-20 & 21.55 & 16.96 (This work)  & 4.59 \\  
    \cdashline{1-4}
    \textbf{$\Delta M$} & 0.05 & -0.65 & - \\ 
    \bottomrule
    \tabnotetext{a}{Comparison between the SQM values obtained from previous works (SQM LIT) and SQM GAMBON model (SQM GMB). For more details, see Section \ref{Sec_GAMBONS}.}
  \end{tabular}
\end{table}

\begin{figure*}[!t]
    \centering
    \includegraphics[width=\textwidth]{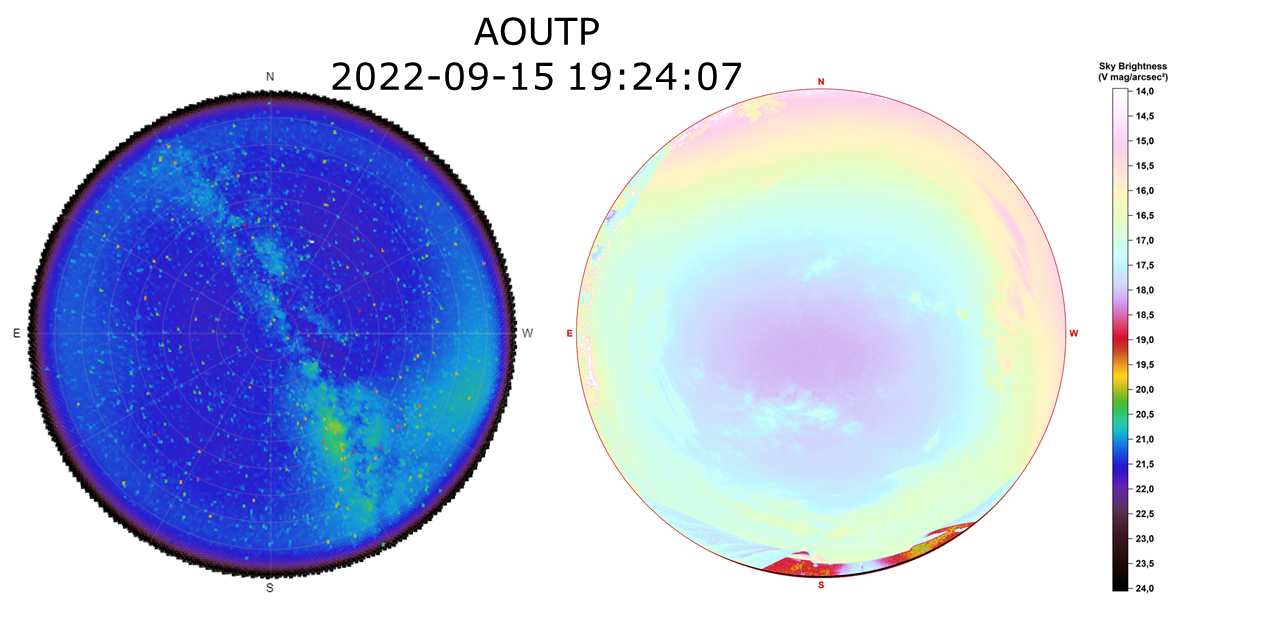}
    \caption{Comparison of the NSB all-sky images generated by the GAMBONS model (left) and the SQC (right) from AOUTP. For more details, see Section \ref{Sec_GAMBONS} and Table \ref{Tab:Gambons}.}
    \label{fig:UTP_Gambons}
\end{figure*}

\begin{figure*}[!t]
    \centering
    \includegraphics[width=\textwidth]{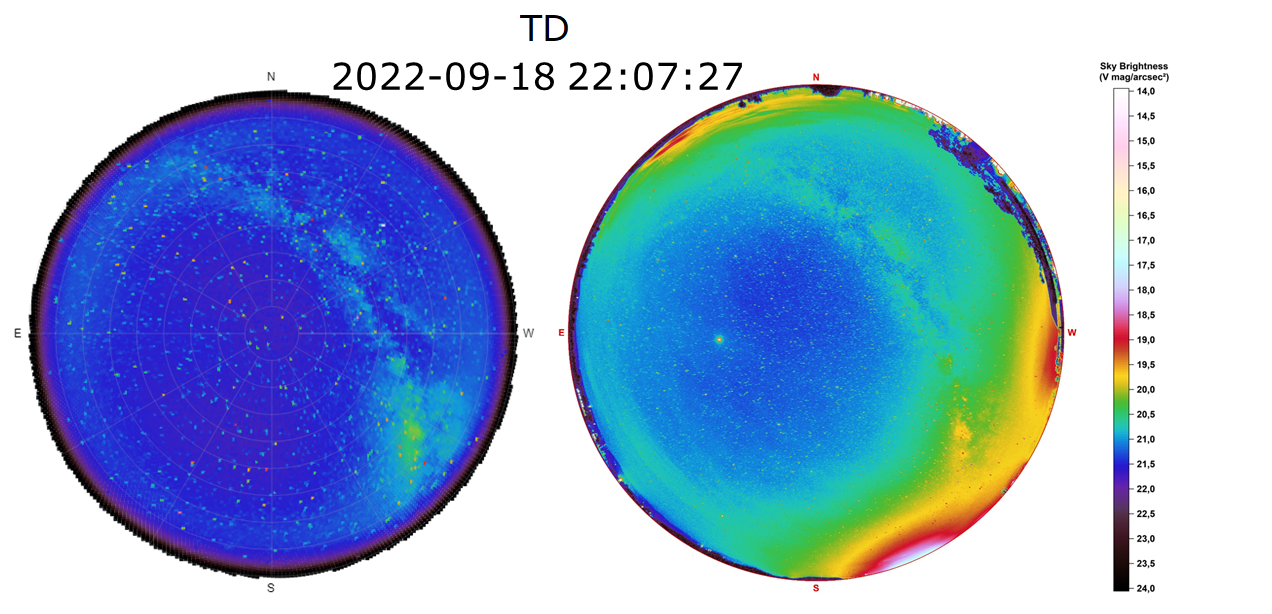}
    \caption{As in Figure \ref{fig:UTP_Gambons}, but for TD. For more details, see Section \ref{Sec_GAMBONS} and Table \ref{Tab:Gambons}.}
    \label{fig:DT_Gambons}
\end{figure*}

\begin{figure*}[!t]
    \centering
    \includegraphics[width=\textwidth]{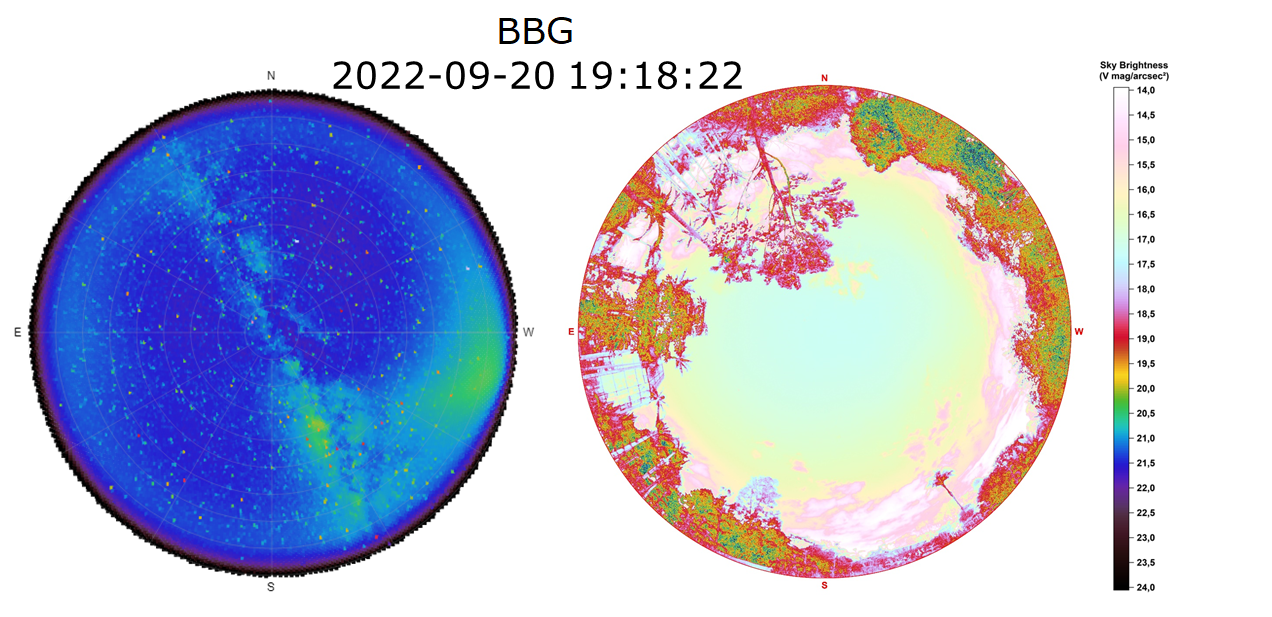}
    \caption{As in Figure \ref{fig:UTP_Gambons}, but for BBG. For more details, see Section \ref{Sec_GAMBONS} and Table \ref{Tab:Gambons}.}
    \label{fig:JBB_Gambons}
\end{figure*}

\begin{figure*}[!t]
    \centering
    \includegraphics[width=\textwidth]{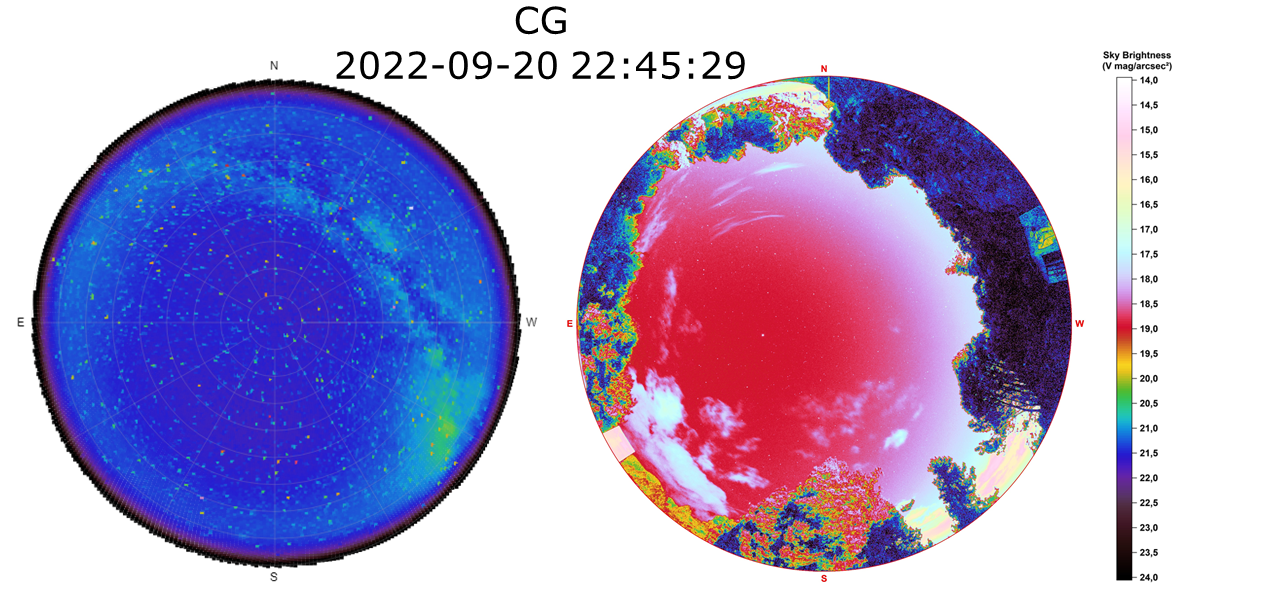}
   \caption{As in Figure \ref{fig:UTP_Gambons}, but for CG. For more details, see Section \ref{Sec_GAMBONS} and Table \ref{Tab:Gambons}.}
    \label{fig:CG_Gambons}
\end{figure*}

Figures \ref{fig:UTP_Gambons}, \ref{fig:DT_Gambons}, \ref{fig:JBB_Gambons}, and \ref{fig:CG_Gambons} display the all-sky images generated by the GAMBONS model, alongside the images obtained by the SQC for each site. For each GAMBONS image, spatial and temporal data were replicated, while for the SQC image, the NSB value was recalculated including the contributions of the resolved stars without applying smoothing. A detailed description of each site is provided below. 

In the AOUTP case, as illustrated in Figure \ref{fig:UTP_Gambons}, 98\% of ALAN in Pereira originates from artificial contributions to NSB. This dominance is evident in Table \ref{Tab:Gambons}, column 7, where the impact of artificial lighting on sky glow is clearly observed. ALAN results from a combination of natural NSB contributions, estimated using the GAMBONS model, and artificial NSB contributions, derived from our data. The overwhelming presence of artificial light highlights the profound transformation of the night sky in urban environments like Pereira. In Figure \ref{fig:DT_Gambons}, it is evident that, with the naked eye, one can still appreciate the stars and our galaxy, the Milky Way in TD. However, the NSB from the cities almost completely dominates the horizon. Table \ref{Tab:Gambons} indicates that approximately 59\% of the sky is dominated by ALAN, which is highly concerning for a Starlight destination whose mission is to conserve and protect the night sky.

Raising awareness among people is crucial, as most inhabitants make their living from tourism, with astrotourism being the main attraction for visitors to the desert \citep{vega2022retos}. Efforts should prioritize Ecopetrol, a private company, and the city of Neiva, as these are the primary sources of light pollution in the desert. This is especially important during lighting transitions, such as the shift to LED technology. While LEDs provide significant economic benefits and enhance conditions for nighttime activities, it is essential to ensure these lights are properly directed toward the ground and avoid over-illumination of spaces. Protecting the night sky is not only vital for environmental conservation but can also boost the local economy by attracting more tourists interested in astrotourism, thereby fostering sustainable development in the region.

Figure \ref{fig:JBB_Gambons} shows that the sky at the BBG is dominated by ALAN. This result was predictable, given that BBG is located within one of the largest metropolises in Latin America. However, this does not imply that the night sky should not be protected. In this particular case, studies are recommended to evaluate how ALAN is affecting the plants in the various ecosystems present in the garden \citep{vargas2019plantas}. Additionally, measures should be taken to protect those collections and plants that are highly sensitive to light changes. 

Moreover, the sky over CG (see Figure \ref{fig:CG_Gambons}), like that of the BBG, is dominated by ALAN due to its proximity to the city of Bogot\'a. However, a 5\% decrease in ALAN is observed compared to BBG. This is because CG is 700 meters higher in elevation and almost 10 km away from the urban center. This difference in altitude and distance allows for better night sky quality in CG, reducing light interference. Additionally, the area's vegetation and natural relief help to mitigate light pollution. For these reasons, CG could be considered a strategic point for astronomical observation and ecological studies. In summary, the analysis of NSB using the GAMBONS model highlights the significant impact of ALAN on astronomical observations and ecological systems. Locations like the TD and BBG demonstrate varying levels of light pollution, emphasizing the need for targeted measures.

\section{Conclusions}\label{Sec_5}
We have presented the first quantitative study of sky quality from different sites in Colombia, one of the most biodiverse countries globally. Our data update the existing values for the zenith in the BBG and TD and provide the first records for Pereira and Cerro Guadalupe. Additionally, the measurements were made with an all-sky instrument, allowing us to obtain information from the entire celestial vault to monitor and quantify the effects of light pollution.

We identified the cities that contribute most to light pollution at each site, along with the azimuths and distances from the observing site. The most polluted site was the BBG, located within Bogot\'a, one of the most populated cities in Latin America. AOUTP is the second most polluted site, located within the city of Pereira, the sixteenth most populated city in Colombia. CG is the third most contaminated site; although it is 700 meters above the city and partially protected by the trees on the hill, it still suffers from significant light pollution. The TD is the least polluted, with an ALAN contribution of 59\%. Measures must be taken to reduce light pollution in these critical areas. The implementation of more efficient and less intrusive lighting technologies can help mitigate this problem.

The Starlight or Dark Sky certification not only ensures that a location offers exceptional conditions for astronomical observation but also serves as a vital tool for protecting and conserving the night sky amidst the growing threat of light pollution. Preserving and renewing these certifications is crucial for safeguarding both the natural environment and the biodiversity that depends on the cycles of light and darkness, as well as for supporting scientific research. Moreover, such certifications stimulate astrotourism, an increasingly important source of income in many regions, promoting sustainable tourism centered on education and the appreciation of nature. To protect the future of these destinations, local administrations need to implement effective measures to conserve dark skies, regularly renew certifications, and foster awareness among tour operators and communities about the significance of environmental protection and sustainability.

In this context, the TD has emerged as Colombia's premier sky observation destination, even before receiving its Starlight certification. This has enabled many local residents to benefit from astrotourism as a valuable source of income. However, local administrations have yet to establish concrete strategies for the conservation and protection of this crucial asset. A study by \citet{vega2022retos} underscores the challenges faced by the Villavieja administration regarding the desert. During our data collection, we were struck by the prevalence of lights directed toward the sky by local businesses and the lack of environmental awareness among tour operators. We hope that our findings can contribute to the development of strategies by both the municipality and government to preserve this unique Starlight destination in Colombia.

On the other hand, astrotourism has become a vital component of sustainable tourism by generating economic benefits for local communities while encouraging environmental conservation. However, this activity must be accompanied by education and awareness initiatives aimed at protecting the natural environment, especially the night sky. Regions like the Tatacoa Desert can capitalize on their dark skies to attract visitors interested in both astronomy and natural beauty. Preserving the night sky is essential to maintaining the desert's irreplaceable heritage. Educating and raising awareness among the local population about the importance of protecting this natural resource is imperative to ensuring that future generations can continue to marvel at the splendor of the starry night sky \citep{vierdayanti2024promoting}.

LED lights represent the new generation of lighting, and it is crucial to pay attention to this transformation in our society. This change in the spectral distribution of the sky will have impacts on fauna, flora, and humans. LED lights were found at all sites analyzed, but their presence is especially notable at TD. In Figure \ref{fig:sqc_DT}, a source of white-blue light can be observed in the RGB map towards the south, and when contrasted with the CCT map, it is evident how this light mixes with the natural brightness of the sky. This phenomenon indicates a degradation in sky quality, resulting in the inability to perceive objects in deep space, the Milky Way, and even stars.

Clouds play a crucial role in the propagation and impact of light pollution, as they reflect and scatter artificial light toward the ground, amplifying its effects and degrading the quality of the night sky, making it difficult to observe celestial objects \citep{kyba2011cloud, jechow2017imaging}. In Colombia, where skies are covered by clouds most of the year due to their geographical location, all-sky measurements are essential to estimate the state of the sky and visualize cloudy areas, thus understanding their interaction with ALAN. This understanding is vital for developing effective light pollution mitigation strategies and ensuring better sky quality for both ecological and astronomical purposes. Therefore, prolonged studies, conducted over months or years, are suggested to more accurately characterize the interaction between clouds and light pollution, implementing measures that protect the night sky and promote sustainable astronomical observation. By enhancing our knowledge in this area, we can better advocate for preserving the night sky for future generations.

Conducting studies on ALAN is crucial for understanding its impact on biodiversity. ALAN alters the natural behaviors of many species, including their reproductive, foraging, and migration patterns \citep{davies2014nature, holker202111, mayer2021light}. This disruption can lead to significant population declines and affect entire ecosystems. Additionally, light pollution influences plants, since artificial light can interfere with their growth and flowering cycles, which in turn impacts the species that depend on them \citep{singhal2019eco, meravi2020effect}. By quantifying light pollution and identifying its sources, we can develop strategies to mitigate its effects, helping to preserve the rich biodiversity found in Colombia, and more specifically in the BBG. These studies are essential for creating policies that balance the needs of human development with the conservation of natural habitats.

Our data represent an initial step in understanding the ALAN phenomenon and will serve as a foundation for future studies on its evolution. All the sites studied are either ecologically important or in proximity to protected areas. The presence of endemic fauna in these locations allows for studies of numerous animal and plant species, making them areas of significant conservation and preservation value.

In future research, we plan to monitor the sites studied in this work to investigate possible seasonal or annual changes in the quality of the night skies. Additionally, we intend to expand our study to include more sites of astronomical and ecological interest. This will enable us to obtain a more comprehensive understanding of the evolution of light pollution and its effects on different ecosystems. We will also seek collaborations with local and regional entities to raise awareness about the importance of preserving the night sky and reducing light pollution in these areas.

The authors would like to thank the referee for helpful comments and suggestions that improved the manuscript. The authors also express their gratitude for all the logistical management carried out by the directors of each site: Dr. Edwin Andr\'es Quintero at AOUTP, Professor Javier Rua at TD, and Mar\'ia Liliana Perdomo at BBG. They also thank all the support received in the selection of the sites from the technical staff at each location. JPUT acknowledges the financial support of the National Agency for Research and Development (ANID) Scholarship Program/Doctorado Nacional/2021-21210732 and the PhD Program in Astronomy at the University of La Serena. M.J.A. acknowledges the financial support of DIDULS/ULS through project PR2353855.


\begin{thebibliography}

\bibitem[Aceituno et al.(2011)]{aceituno2011all}
Aceituno, J., S{\'a}nchez, S.~F. and Aceituno, F.~J., et al. 2011, \pasp, 123, 907, https://doi.org/10.1086/661918

\bibitem[Aguilera Camargo (2012)]{aguilera2012evaluacion}
Aguilera Camargo, J.~N. \ 2012, Repositorio - Uniandes, http://hdl.handle.net/1992/25111

\bibitem[Alarcon et al.(2021)]{alarcon2021natural}
Alarcon, M.~R., Serra-Ricart, M, Lemes-Perera, S., et al. 2021, The Astronomical Journal, 162, 1, https://doi.org/10.3847/1538-3881/abfdaa

\bibitem[Andrade-C (2011)]{andrade2011estado}
Andrade-C, M.~G. \ 2011, Revista De La Academia Colombiana De Ciencias Exactas, Físicas Y Naturales, 35, 137, https://doi.org/10.18257/raccefyn.35(137).2011.2424

\bibitem[Angeloni et al.(2024)]{angeloni2024toward}
Angeloni, R., Uchima-Tamayo, J.~P., Jaque Arancibia, M., et al. 2024, The Astronomical Journal, 167, 2, https://doi.org/10.3847/1538-3881/ad165c

\bibitem[Arbel{\'a}ez-Cardona et al.(2020)]{arbelaez2020estimating}
Arbel{\'a}ez-Cardona, D., Silva-Villa, E., Galvez, J. et al. 2020, International Journal of Remote Sensing, 41, 14, https://doi.org/10.1080/01431161.2020.1727051

\bibitem[Aub{\'e} et al.(2014)]{aube2014evaluation}
Aub{\'e}, M., Fortin, N., Turcotte, S. et al. 2014, \pasp, 126, 945, https://doi.org/10.1086/679227

\bibitem[Aub{\'e} et al.(2016)]{aube2016spectral}
Aub{\'e}, M., Kocifaj, M., Zamorano, J. et al. 2016, Journal of Quantitative Spectroscopy and Radiative Transfer, 181, https://doi.org/10.1016/j.jqsrt.2016.01.032

\bibitem[Bar\'a et al.(2019)]{bara2019absolute}
Bar\'a, S., Tapia, C.~E., Zamorano, J. et al. 2019, Sensors, 19, 6, https://doi.org/10.3390/s19061336

\bibitem[Bar\'a \& Falchi (2023)]{bara2023artificial}
Bar\'a, S. \& Falchi, F. \ 2023, Philosophical Transactions of the Royal Society B, 378, 1892, https://doi.org/10.1098/rstb.2022.0352

\bibitem[Barentine (2022)]{barentine2022night}
Barentine, J.~C. 2022, Nature Astronomy, 6, 10, https://doi.org/10.1038/s41550-022-01756-2

\bibitem[Ben{\'i}tez Garc{\'i}a (2016)]{benitez2016eficacia}
Ben{\'i}tez Garc{\'i}a, L.~M. \ 2016, Repositorio - Universidad Libre, https://hdl.handle.net/10901/10198

\bibitem[Bertolo et al.(2019)]{bertolo2019measurements}
Bertolo, A., Binotto, R., Ortolani, S. et al. 2019, Journal of Imaging, 5, 5,
https://doi.org/10.3390/jimaging5050056

\bibitem[Boh\'orquez-Alfonso (2008)]{bohorquez2008arriba}
Boh\'orquez-Alfonso, I.~A. 2008, Cuadernos De Vivienda Y Urbanismo, 1, 1, https://revistas.javeriana.edu.co/index.php/cvyu/article/view/5485

\bibitem[Bonilla Romero (2011)]{romero2011aproximaciones}
Bonilla Romero, J. 2011, Revista de Topografía AZIMUT, 3, 
https://revistas.udistrital.edu.co/index.php/azimut/article/view/4055

\bibitem[Bonilla Romero et al.(2017)]{romeroarqueoastronomia}
Bonilla Romero, J., Bustos velazco, E.~H. \& Reyes, J.~D. 2017, Revista Cient{\'\i}fica, Numero Especial,  https://doi.org/10.14483/udistrital.jour.RC.2017.27.a15

\bibitem[Cadena-Vargas et al.(2020)]{vargas2019plantas}
Cadena-Vargas, C.~E., S{\'a}nchez Callejas, S.~D. \& Morales Pisco, A.~F. 2020, Revista Facultad de Ciencias B{\'a}sicas, 15, 2, https://doi.org/10.18359/rfcb.4382

\bibitem[Cadena-Vargas et al.(2021)]{cadena2021coleccion}
Cadena-Vargas, C.~E., S{\'a}nchez Callejas, S.~D. \& Vel{\'a}squez Ni{\~n}o, J. 2021, Biota colombiana, 22, 2, https://doi.org/10.21068/c2021.v22n02a10 

\bibitem[Cao \& Bai (2014)]{cao2014quantitative}
Cao, C. \& Bai, Y. 2014, Remote sensing, 6, 12, https://doi.org/10.3390/rs61211915

\bibitem[Chang et al.(2012)]{chang2012light}
Chang, Moon-Hwan, Das, Diganta, Varde, P.~V. et al. 2012, Microelectronics Reliability, 52, 5, https://doi.org/10.1016/j.microrel.2011.07.063

\bibitem[Chaparro Molano et al.(2017)]{molano2017low}
Chaparro Molano, G., Ramírez Suárez, O.~L., Restrepo Gaitán, O.~A. et al. 2017, \pasp, 129, 980, https://doi.org/10.1088/1538-3873/aa83fe

\bibitem[Cho et al.(2015)]{cho2015effects}
Cho, Y., Ryu, S.,  Lee, B. R., et al. 2015, Chronobiology International, 32, 9, https://doi.org/10.3109/07420528.2015.1073158

\bibitem[Cho et al.(2017)]{cho2017white}
Cho, J., Park, J.~H., Kim, J.~K. et al. 2017, Laser \& photonics reviews, 11, 2,  https://doi.org/10.1002/lpor.201600147

\bibitem[Cinzano et al.(2000)]{cinzano2000artificial}
Cinzano, P., Falchi, F., Elvidge, C.~D. et al. 2000, \mnras, 318, 3,  https://doi.org/10.1046/j.1365-8711.2000.03562.x

\bibitem[Cinzano (2005)]{cinzano2005night}
Cinzano, P. 2005, ISTIL Internal Report, 9, 1

\bibitem[C-S{\'a}nchez et al.(2019)]{c2019astrotourism}
C-S{\'a}nchez, E., S{\'a}nchez-Medina, A.~J., Alonso-Hern{\'a}ndez, J.~B., et al. 2019, Sensors, 19, 13, https://doi.org/10.3390/s19132840

\bibitem[Davies et al.(2014)]{davies2014nature}
Davies, T.~W., Duffy, J.~P., Bennie, J., et al. 2014, Frontiers in Ecology and the Environment, 12, 6, https://doi.org/10.1890/130281

\bibitem[Deprato et al.(2024)]{deprato2024influence}
Deprato, A., Maidstone, R., Cros, A.~P, et al. 2024, BMC medicine, 22, 67, https://doi.org/10.1186/s12916-024-03291-5

\bibitem[Deverch{\`e}re et al.(2022)]{deverchere2022towards}
Deverch{\`e}re, P., Vauclair, S., Bosch, G., et al. 2022, Scientific Reports, 12, 1, https://doi.org/10.1038/s41598-022-21460-5

\bibitem[Dill et al.(2020)]{dill2020badland}
Dill, H.~G., Andrei, B., Sorin-Ionut, B., et al. 2020, Catena, 194, https://doi.org/10.1016/j.catena.2020.104696

\bibitem[Duriscoe (2016)]{duriscoe2016photometric}
Duriscoe, D.~M. 2016, \jqsrt, 181, https://doi.org/10.1016/j.jqsrt.2016.02.022

\bibitem[Durmus (2022)]{durmus2022correlated}
Durmus, D. 2022, Lighting Research \& Technology, 54, 4, https://doi.org/10.1177/14771535211034330

\bibitem[Falchi et al.(2011)]{falchi2011limiting}
Falchi, F., Cinzano, P., Elvidge, C.~D. et al. 2011, Journal of environmental management, 92, 10, https://doi.org/10.1016/j.jenvman.2011.06.029

\bibitem[Falchi et al.(2016)]{falchi2016new}
Falchi, F., Cinzano, P., Duriscoe, D., et al. 2016, Science advances, 2, 6, https://doi.org/10.1126/sciadv.1600377

\bibitem[Falchi et al.(2021)]{falchi2021computing}
Falchi, F. \& Bar{\'a}, S., 2021, Natural Sciences, 1, 2,  https://doi.org/10.1002/ntls.10019

\bibitem[Falchi et al.(2023)]{falchi2023light}
Falchi, F., Ramos, F., Bar{\'a}, S., et al. 2023, \mnras, 519, 1,  https://doi.org/10.1093/mnras/stac2929

\bibitem[Fl{\'o}rez et al.(2013)]{florez2013paleosuelos}
Fl{\'o}rez, M.~T., Parra, L.~N., Jaramillo, D.~F., et al. 2013, Revista De La Academia Colombiana De Ciencias Exactas, Físicas Y Naturales, 37, 143, https://doi.org/10.18257/raccefyn.6

\bibitem[Gabriel et al.(2017)]{gabriel2017resources}
Gabriel, K.~M.~A., Kuechly, H.~U., Falchi F., et al. 2017, International journal of Biometeorology, 61, https://doi.org/10.1007/s00484-016-1187-y

\bibitem[Gallaway et al.(2010)]{gallaway2010economics}
Gallaway, T., Olsen, R.~N. \& Mitchell, D.~M., 2010, Ecological economics, 69, 3,
https://doi.org/10.1016/j.ecolecon.2009.10.003

\bibitem[Galvis et al.(2019)]{galvis2019space}
Galvis, H.~D., Galeano, D. \& Quintero Salazar, E.~A., 2019, Sun and Geosphere, 14, 2,
https://doi.org/10.31401/SunGeo.2019.02.10

\bibitem[Garc{\'\i}a Sierra \& Ram{\'\i}rez Cardona (2011)]{garcia2011estudio}
Garc{\'\i}a Sierra, J.~H. \& Ram{\'\i}rez Cardona, J.~L.,  2011, Luna Azul, 32, 
 https://revistasojs.ucaldas.edu.co/index.php/lunazul/article/view/1242

 \bibitem[Garz{\'o}n D{\'\i}az(2014)]{garzon2014educacion}
Garz{\'o}n D{\'\i}az, F.~A.,  2014, Revista latinoamericana de bio{\'e}tica, 14, 1,  https://doi.org/10.18359/rlbi.498

\bibitem[Gaston et al.(2013)]{gaston2013ecological}
Gaston, K.~J, Bennie, J., Davies, T.~W. et al. 2013, Biological reviews, 88, 4,  https://doi.org/10.1111/brv.12036

\bibitem[G{\'o}ez Ther{\'a}n \& Vargas Dom{\'\i}nguez (2021)]{goez2021comparative}
G{\'o}ez Ther{\'a}n, C. \& Vargas Dom{\'\i}nguez, S. \ 2021, RMxAA, 57, 1,
https://doi.org/10.22201/ia.01851101p.2021.57.01.03

\bibitem[H{\"a}nel et al.(2018)]{hanel2018measuring}
H{\"a}nel, A., Posch, T., Ribas, S.~J. et al. 2018, \jqsrt, 205, https://doi.org/10.1016/j.jqsrt.2017.09.008

\bibitem[Henderson (2010)]{henderson2010valuing}
Henderson, D., 2010, Environmental Philosophy, 7, 1, https://www.jstor.org/stable/26168027

\bibitem[H{\"o}lker et al.(2010)]{holker2010dark}
H{\"o}lker, F., Moss, T., Griefahn, B., et al. 2010, Ecology and Society, 15, 4, https://www.jstor.org/stable/26268230

\bibitem[H{\"o}lker et al.(2021)]{holker202111}
H{\"o}lker, F., Bolliger, J., Davies, T., et al. 2021, Frontiers in Ecology and Evolution, 9,
https://doi.org/10.3389/fevo.2021.767177

\bibitem[Hung et al.(2021)]{hung2021changes}
Hung, Li-Wei, Anderson, S.~J., Pipkin, A. et al. 2021, Journal of Environmental Management, 292, https://doi.org/10.1016/j.jenvman.2021.112776

\bibitem[Iglesias et al.(2023)]{iglesias2023daytime}
Iglesias, F.~A., Francile, C., Lazarte-Gelmetti, J., et al. 2023, Solar Physics, 298, 46,  https://doi.org/10.1007/s11207-023-02139-0

\bibitem[Jechow et al.(2017)]{jechow2017imaging}
Jechow, A.,  Koll{\'a}th, Z., Ribas, S.~J., et al. 2017, Scientific Reports, 7, 1,  https://doi.org/10.1038/s41598-017-06998-z

\bibitem[Jechow et al.(2019)]{jechow2019using}
Jechow, A., H{\"o}lker, F. \& Kyba, C.~C.~M. \ 2019, Scientific reports, 9, 1,  https://doi.org/10.1038/s41598-018-37817-8

\bibitem[Jim{\'e}nez-Alvarado et al.(2017)]{jimenez2017ciudades}
Jim{\'e}nez-Alvarado, J.~S., Moreno-D{\'\i}az, C., Alfonso, A.~F. et al. 2017, Mammalogy Notes, 4, 1,  https://doi.org/10.47603/manovol4n1.37-41

\bibitem[Jim{\'e}nez Villariaga et al.(2017)]{villariaga2017obtaining}
Jim{\'e}nez Villariaga, S., Quintero Salazar, E.~A., \& Aguirre Galvis, J.~A., 2017, TECCIENCIA, 12, 23, https://doi.org/10.18180/tecciencia.2017.23.6 

\bibitem[Koll{\'a}th (2010)]{kollath2010measuring}
Koll{\'a}th, Z. 2010, Journal of Physics: Conference Series, 218, 1, https://doi.org/10.1088/1742-6596/218/1/012001

\bibitem[Koll{\'a}th et al.(2020)]{kollath2020introducing}
Koll{\'a}th, Z., Cool, A., Jechow, A. et al. 2020,\jqsrt, 253, https://doi.org/10.1016/j.jqsrt.2020.107162

\bibitem[Krisciunas et al.(2007)]{krisciunas2007optical}
Krisciunas, K., Semler, D.~R, Richards, J. et al. 2007, \pasp, 119, 856,  https://doi.org/10.1086/519564

\bibitem[Kyba et al.(2011)]{kyba2011cloud}
Kyba, C.~C.~M., Ruhtz, T., Fischer, J. et al. 2011, PloS one, 6, 3, https://doi.org/10.1371/journal.pone.0017307

\bibitem[Kyba et al.(2017)]{kyba2017artificially}
Kyba, C.~C.~M., Kuester, T., S{\'a}nchez de Miguel, A. et al. 2017, Science advances, 3, 11, https://doi.org/10.1126/sciadv.1701528

\bibitem[Kyba et al.(2023)]{kyba2023}
Kyba, C.~C.~M., Altintas, Y.~O., Walker, C.~E. et al. 2023, Science, 379, 6629, https://doi.org/10.1126/science.abq7781

\bibitem[Longcore \& Rich (2004)]{longcore2004ecological}
Longcore, T. \& Rich, C. \ 2004, Frontiers in Ecology and the Environment, 2, 4,
https://doi.org/10.1890/1540-9295(2004)002[0191:ELP]2.0.CO;2

\bibitem[Luo et al.(2023)]{luo2023effects}
Luo, W., Kramer, R., Kompier, M. et al. 2023, Building and Environment, 231,  https://doi.org/10.1016/j.buildenv.2022.109944

\bibitem[Mander et al.(2023)]{mander2023measure}
Mander, S., Alam, F., Lovreglio, R., et al. 2023, Sustainable Cities and Society, 92 https://doi.org/10.1016/j.scs.2023.104465

\bibitem[Mar\'in G\'omez (2022)]{marin2022artificial}
Mar\'in G\'omez, O.~H. \ 2022, Animals, 12, 8, https://doi.org/10.3390/ani12081015

\bibitem[Masana et al.(2021)]{masana2021multiband}
Masana, E., Carrasco, J.~M., Bar{\'a}, S., et al. 2021, \mnras, 501,4, https://doi.org/10.1093/mnras/staa4005

\bibitem[Masana et al.(2022)]{masana2022enhanced}
Masana, E., Bar{\'a}, S., Carrasco, J.~M. et al. 2022, International Journal of Sustainable Lighting, 24, 1, https://doi.org/10.26607/ijsl.v24i1.119

\bibitem[Mayer-Pinto et al.(2022)]{mayer2021light}
Mayer-Pinto, M., Jones, T.~M, Swearer, S.~E. et al. 2022, UCL Open Environment, 4,  https://doi.org/10.14324/111.444/000103.v1

\bibitem[Mej{\'\i}a (2006)]{del2006monserrate}
 Mej{\'\i}a, M.~P. 2006, Fronteras de la Historia, 11, https://www.redalyc.org/articulo.oa?id=83301108

 \bibitem[Meravi \& Kumar Prajapati (2018)]{meravi2020effect}
Meravi, N. \& Kumar Prajapati, S., 2018, Biological Rhythm Research, 51, 1,
https://doi.org/10.1080/09291016.2018.1518206

 \bibitem[Meza (2008)]{meza2008urbanizacion}
Meza, C.~A. 2008, Revista colombiana de antropolog{\'\i}a, 44, 2,  https://www.redalyc.org/articulo.oa?id=105012451007

\bibitem[Mills et al.(2007)]{mills2007effect}
Mills, P.~R, Tomkins, S.~C. \& Schlangen, L.~J. \ 2007, Journal of circadian rhythms, 5, https://doi.org/10.1186/1740-3391-5-2

\bibitem[Mitchell \& Gallaway (2019)]{mitchell2019dark}
Mitchell, D.~M. \& Gallaway, T., 2019, Tourism Review, 74, 4,
https://doi.org/10.1108/TR-10-2018-0146

\bibitem[Montealegre Vega \& Garavito Gonz\'alez (2022)]{vega2022retos}
Montealegre Vega, A. \& Garavito Gonz\'alez, L. \ 2022, TURPADE. Turismo, Patrimonio y Desarrollo, 1, 17, https://revistaturpade.lasallebajio.edu.mx/index.php/turpade/article/view/65

\bibitem[Montes et al.(2021)]{montes2021middle}
Montes, C., Silva, C.~A., Bayona, G.~A., et al. 2021, Frontiers in Earth Science, 8, https://doi.org/10.3389/feart.2020.587022

\bibitem[M{\"u}ller et al.(2011)]{muller2011measuring}
M{\"u}ller, A., Wuchterl, G. \&  Sarazin, M. \ 2011, Revista Mexicana de Astronom{\'\i}a y Astrof{\'\i}sica, 41, https://www.redalyc.org/articulo.oa?id=57120784014

\bibitem[Nair \& Dhoble (2015)]{nair2015perspective}
Nair, G.~B. \& Dhoble, S.~J., 2015, Luminescence, 30, 8,
https://doi.org/10.1002/bio.2919

\bibitem[Plauchu-Frayn et al.(2017)]{plauchu2017night}
Plauchu-Frayn, I., Richer, M.~G., Colorado, E., et al. 2017, \pasp, 129, 973, https://doi.org/10.1088/1538-3873/129/973/035003

\bibitem[Posh et al.(2018)]{posch2018systematic}
Posh, T., Binder, F. \&  Puschnig, J. \ 2018, \jqsrt, 211, https://doi.org/10.1016/j.jqsrt.2018.03.010

\bibitem[Poveda et al.(2011)]{poveda2011hydro}
Poveda, G., \'Alvarez, D.~M. \&  Rueda,  O.~A. \ 2011, Climate Dynamics, 36, https://doi.org/10.1007/s00382-010-0931-y

\bibitem[Riegel (1973)]{riegel1973light}
Riegel, K.~W., 1973, Science, 179, 4080, https://doi.org/10.1126/science.179.4080.1285

\bibitem[Rodr\'iguez Ramos (2015)]{rodriguez2015plan}
Rodr\'iguez Ramos, N.~B. 2015, Repositorio Universidad Javeriana, http://hdl.handle.net/10554/17958

\bibitem[Rueda-Espinosa et al.(2023)]{rueda2023illuminating}
Rueda-Espinosa, K.~J, Guerrero-Guio, A.~F., Vargas-Dominguez, S., et al. 2023, Revista de la Academia Colombiana de Ciencias Exactas, F{\'\i}sicas y Naturales, 47, 183, https://doi.org/10.18257/raccefyn.1867

\bibitem[Rueda Esteban (2017)]{esteban2017gestion}
Rueda Esteban, N.~R, 2017, PASOS Revista de Turismo y Patrimonio Cultural, 15, 1, https://www.pasosonline.org/Publicados/15117/PASOS51.pdf\#page=87

\bibitem[S{\'a}nchez-Gonz{\'a}lez et al.(2020)]{sanchez2021urbanization}
S{\'a}nchez-Gonz{\'a}lez, K., Aguirre-Obando, O. \& R{\'\i}os-Chel{\'e}n, A.~A. 2020, Ethology Ecology \& Evolution, 33, 4, https://doi.org/10.1080/03949370.2020.1837963 

\bibitem[Singhal et al.(2019)]{singhal2019eco}
Singhal, R.~K., Kumar, M. \& Bose, B. 2019, Russian Journal of Plant Physiology, 66, 2, https://doi.org/10.1134/S1021443719020134

\bibitem[Svechkina et al.(2020)]{svechkina2020impact}
Svechkina, A., Portnov, B.~A. \& Trop, T.  2020, Landscape Ecology, 35, 8, 
https://doi.org/10.1007/s10980-020-01053-1

\bibitem[Tsao et al.(2010)]{tsao2010world}
Tsao, J., Saunders, H.~D., Creighton,J.~R.  et al. 2010, Journal of Physics D: Applied Physics, 43, 35, https://doi.org/10.1088/0022-3727/43/35/354001

\bibitem[Tovmassian et al.(2016)]{tovmassian2016astroclimatic}
Tovmassian, G., Hernandez, M., Ochoa,J.~L.  et al. 2016, \pasp, 128, 961, https://doi.org/10.1088/1538-3873/128/961/035004

\bibitem[Urrego Guevara (2016)]{urrego2016analisis}
Urrego Guevara, G.~A. \ 2016, Repositorio - Universidad Libre, https://hdl.handle.net/10901/8897

\bibitem[Varela Perez (2023)]{varela2023increasing}
Varela Perez, A.~M. \ 2023, Science, 380, 6650, https://doi.org/10.1126/science.adg0269

\bibitem[Verheijen (1985)]{verheijen1985photopollution}
Verheijen, F.~J. \ 1985, Experimental biology, 44, 1

\bibitem[Vierdayanti et al.(2024)]{vierdayanti2024promoting}
Vierdayanti, K., Kunjaya, C., Herdiwijaya, D. et al. 2024, Journal of Physics: Conference Series, 1, https://doi.org/10.1088/1742-6596/2773/1/012020

\bibitem[Walker (1970)]{walker1973light}
Walker, M.~F., 1970, \pasp, 82, 672, https://doi.org/10.1086/128945

\bibitem[Zamorano et al.(2016)]{zamorano2016stars4all}
Zamorano, J., Garc{\'\i}a, C., Tapia, C. et al. 2016, International Journal of Sustainable Lighting, 18, 1, https://doi.org/10.26607/ijsl.v18i0.21

\bibitem[Zielinska-Dabkowska (2019)]{zielinska2019urban}
Zielinska-Dabkowska, K.~M. \ 2019, Urban Lighting for People, https://doi.org/10.4324/9780367814588-2

\bibitem[Zielinska-Dabkowska et al.(2020)]{zielinska2020assessment}
Zielinska-Dabkowska, K.~M., Xavia, K. \& Bobkowska, K. \ 2020, Sustainability, 12, 12, 
https://doi.org/10.3390/su12124997



\end{thebibliography}
\end{document}